\documentclass[conference]{IEEEtran}
\usepackage{epsfig,endnotes}

\newcommand{\sect}{Sect.~}
\newcommand{\etal}{\emph{et~al.}\xspace}
\newcommand{\ournameNoSpace}{\mbox{AuthentiSense}}
\newcommand{\ourname}{\ournameNoSpace\xspace}

\newcommand{\nComparison}{n-shot\xspace}

\newcommand{\fscoreNoSpace}{F1-Score}
\newcommand{\fscore}{\fscoreNoSpace\xspace}
\newcommand{\fscores}{\fscoreNoSpace s\xspace}
\newcommand{\authentWindowLength}{Authentication window length\xspace}
\usepackage{algpseudocode}
\usepackage{amsmath}
\usepackage{amsthm}
\usepackage{amsfonts}
\usepackage{amssymb}
\usepackage{array}
\usepackage{booktabs}
\usepackage{boxedminipage}
\usepackage{calrsfs}
\usepackage{caption}
\usepackage{color}
\usepackage{fancyhdr}
\usepackage{float}
\usepackage{graphicx}
\usepackage{hyperref}
\hypersetup{
    colorlinks=false,
    pdfborder={0 0 0},
}
\usepackage{listings}
\usepackage{multirow}
\usepackage{nicefrac}
\usepackage{pifont}
\usepackage{sidecap}
\usepackage{subcaption}
\usepackage{hyperref}
\usepackage{url}
\usepackage{verbatim}
\usepackage{xspace}
\usepackage[ff,sets,adversary,keys,primitives,operators,lambda,logic,notions]{style/cryptocode}
\usepackage[font=normalsize,labelfont=bf]{caption}
\usepackage{makecell}
\renewcommand\theadfont{\small\sffamily}

\lstdefinelanguage{javascript}{
  keywords={typeof, new, true, false, catch, function, return, null, catch, switch, var, if, in, while, do, else, case, break},
  keywordstyle=\color{blue}\bfseries,
  ndkeywords={class, export, boolean, throw, implements, import, this},
  ndkeywordstyle=\color{black}\bfseries,
  identifierstyle=\color{black},
  sensitive=false,
  comment=[l]{//},
  morecomment=[s]{/*}{*/},
  commentstyle=\color{purple}\ttfamily,
  stringstyle=\color{red}\ttfamily,
  morestring=[b]',
  morestring=[b]"
}

\newcommand{\squishlist}{\begin{itemize}[itemsep=0pt,parsep=0pt,topsep=0pt,partopsep=0pt,leftmargin=1em,labelwidth=1em,labelsep=0.5em]}
\newcommand{\squishlistend}{\end{itemize}}
\newcommand{\squishend}{\end{itemize}}
\newcommand{\squishenum}{\begin{enumerate}[itemsep=0.5pt,parsep=0pt,topsep=0pt,partopsep=0pt,leftmargin=1.5em,labelwidth=1em,labelsep=0.5em]{}}
\newcommand{\squishenumend}{\end{enumerate}}

\newcommand\myurl[2]{\url{#1}}

\newcommand{\captionfonts}{\small}
\makeatletter  
\long\def\@makecaption#1#2{%
  \vskip 0.1in
  \sbox\@tempboxa{{\captionfonts #1: #2}}%
  \ifdim \wd\@tempboxa >\hsize
    {\captionfonts #1: #2\par}
  \else
    \hbox to\hsize{\hfil\box\@tempboxa\hfil}%
  \fi
  \vskip 0in}
\makeatother   

\definecolor{Mycolor}{HTML}{009B55}

\DeclareUnicodeCharacter{65533}{{\"O}}
\newcommand{\STAB}[1]{\begin{tabular}{@{}c@{}}#1\end{tabular}}
\ifCLASSINFOpdf
\else
\fi
\begin{document}
%
\title{\ourname: A Scalable Behavioral Biometrics Authentication Scheme using Few-Shot Learning \\ for Mobile Platforms}

{\author
{\IEEEauthorblockN{Hossein Fereidooni\IEEEauthorrefmark{1},
Jan König\IEEEauthorrefmark{2},
Phillip Rieger\IEEEauthorrefmark{1},
Marco Chilese\IEEEauthorrefmark{1},
Bora G\"{o}kbakan\IEEEauthorrefmark{3}}
 Moritz Finke\IEEEauthorrefmark{2},
Alexandra Dmitrienko\IEEEauthorrefmark{2}, and
Ahmad-Reza Sadeghi\IEEEauthorrefmark{1}
\IEEEauthorblockA{\IEEEauthorrefmark{1}Technical University of Darmstadt, Germany\\
}
\IEEEauthorblockA{\IEEEauthorrefmark{2}University of W\"urzburg, Germany}
\IEEEauthorblockA{\IEEEauthorrefmark{3}KOBIL, Germany}
 \\[4.0ex]
}}


%


\IEEEoverridecommandlockouts
\makeatletter\def\@IEEEpubidpullup{6.5\baselineskip}\makeatother
\IEEEpubid{\parbox{\columnwidth}{
Network and Distributed System Security (NDSS) Symposium 2023\\
   27 February - 3 March 2023, San Diego, CA, USA\\
   ISBN 1-891562-83-5\\
   https://dx.doi.org/10.14722/ndss.2023.23194\\
   www.ndss-symposium.org
}
\hspace{\columnsep}\makebox[\columnwidth]{}}

\maketitle

\begin{abstract}
Mobile applications are widely used for online services sharing a large amount of personal data online. One-time authentication techniques such as passwords and physiological biometrics (e.g., fingerprint, face, and iris) have their own advantages but also disadvantages since they can be stolen or emulated, and do not prevent access to the underlying device, once it is unlocked. To address these challenges, complementary authentication systems based on behavioural biometrics have emerged. The goal is to continuously profile users based on their interaction with the mobile device. However, existing behavioural authentication schemes are not (i) user-agnostic meaning that they cannot dynamically handle changes in the user-base without model re-training, or (ii) do not scale well to authenticate millions of users. \\
In this paper, we present \ourname, a user-agnostic, scalable, and efficient behavioural biometrics authentication system that enables continuous authentication and utilizes only motion patterns (i.e., accelerometer, gyroscope and magnetometer data) while users interact with mobile apps. Our approach requires neither manually engineered features nor a significant amount of data for model training. We leverage a few-shot learning technique, called Siamese network, to authenticate users at a large scale. We perform a systematic measurement study and report the impact of the parameters such as interaction time needed for authentication and n-shot verification (comparison with enrollment samples) at the recognition stage. Remarkably, \ourname achieves high accuracy of up to 97\% in terms of F1-score even when evaluated in a few-shot fashion that requires only a few behaviour samples per user (3 shots). Our approach accurately authenticates users only after 1 second of user interaction. For \ourname, we report a FAR and FRR of 0.023 and 0.057, respectively.
\end{abstract}

\textbf{Keywords -- Behavioural Biometrics, Authentication, Few-shot Learning, and Siamese Networks.} 

\section{Introduction}
\label{sec:introduction}

Today, traditional authentication methods such as multi-factor methods (based on SMS, or authenticator apps) are not sufficiently robust to prevent sophisticated  attacks~\cite{akhtar} leaving a gap for persistent, adaptive and user-friendly  authentication schemes. One-time authentication methods, such as passwords or physiological biometrics on mobile platforms require users to explicitly interact with their devices, referred to as explicit authentication, to gain access to the device. 

While physiological biometrics promise to create a safer and more convenient alternative to passwords, they are not infallible and suffer from inherent disadvantages related to the nature of physical characteristics. For instance, once the physical characteristics are exposed, they can be reused maliciously multiple times~\cite{Czajka,Shelton}. On the other hand, behavioral biometrics authentication systems aim to address this challenge through an additional layer of security which frequently and unobtrusively monitors the user's interaction with the device ~\cite{Revett}. The approach is to identify unique individual regularities in user's behavior and analyze several parameters such as touch gestures, navigation, and motion patterns, during user's online activities, to detect potential irregularities not related to the real user. Behavioral authentication systems promise to provide increased security due to the dynamic authentication and the resilience to circumvention  their traits are hard to emulate or copy~\cite{Elmiligi22}.

An advantage of behavioral biometrics is that they can be collected in non-obtrusively without disturbing the normal service utilization. Moreover, they enable constant user monitoring and ensure that only the authorized user can use the system, even after an initial identity check has been performed. This ensures a frictionless authentication process and prevents identity fraud, account takeover, and automated attacks such as  recognizing non-human device activity (bots), Remote Access Trojans and emulators~\cite{arvato}. Behavioral biometrics authentication are rapidly emerging and being deployed by major enterprises, such as Mastercard and Deutsche Bank~\cite{mastercard,deutschebank}.

Various behavioral authentication approaches for users of mobile devices have been proposed in the literature so far: they collect user's unique individual features based on, e.g., motion sensors~\cite{buriro2016hold,li2020scanet,hu2018cnnauth,lee2017secure,buriro2017please,Centeno2018}, touch gestures ~\cite{lu2015safeguard,xu2014towards,frank2012touchalytics,zhao2013continuous}, or their combinations~\cite{deb2019actions,bo2013silentsense,buriro2015touchstroke,buriro2017behavioral,zhu2013sensec}, and detect irregularities during an entire online session. In this paper we focus on behavioral authentication solutions which utilize motion sensor values.

Despite their added value, existing behavioural authentication solutions still face several challenges: i) they often require a large amount of training data to build an accurate model~\cite{neverova},  ii) they are not scalable and only work for a limited number of users they were trained for (not user-agnostic)~\cite{shen2016},
iii) need a model per user to improve model performance which could result in a resource intensive implementation in deployment phase~\cite{Centeno2018}, and iv) often require a long interaction time to preciously learn users behaviour~\cite{sitova,Volaka}.

\textbf{Our Goal and Contributions.} We present a framework for continuous user authentication that leverages behavioural biometrics and aims at tackling the aforementioned challenges. 
Our scheme is (i) efficient and does not require 
hand-crafted features for model training, (ii) scalable to authenticate millions of users, and (iii) user-agnostic, i.e., does not need to be re-trained when users are dynamically changing (i.e., joining or leaving the system).

For this, we utilize well-established few-shot learning technique~\cite{oneshot,Fink2004} in which the model can learn how to perform user authentication with a small amount of data (hence few-shot). 
More specifically, we use the Siamese neural network~\cite{NIPS1993} which can also be used when even only one user behavioral sample is available (one-shot learning). As the name suggests, our model stems from 'Siamese twins'~\cite{quigley} where two networks share weights and biases with the intention to learn similarities as well as dissimilarities between input data (i.e., users behavior).
The Siamese networks have been already utilized for behavioral user authentication~\cite{deb2019actions,Centeno2018}. However, they do not solve the scalability problem (cf.  \sect~\ref{sec:related}). \\
Our contributions are summarized as follows:

\begin{itemize}
    \item We present \ourname, a user-agnostic behavioral authentication system that utilizes few-shot learning to authenticate users quickly at large scale without requiring a significant amount of data for model training. 
    \item We develop an end-to-end neural network architecture that can dynamically handle user changes without requiring re-training and feature engineering of input data. It only  utilizes motion patterns (i.e., accelerometer, gyroscope and magnetometer data). Our approach can achieve an accuracy of 97\% in terms of F1-score for 3-shot verification, and can accurately authenticate users only after 1 second of user interaction. 
    \item  We conduct an extensive and systematic measurement study in which we investigate the impact of different parameters such as training strategy (pairwise or triplet), interaction time needed for authentication, and  n-shot verification (comparison with enrollment samples) at recognition stage in our system.
    
\end{itemize}

\section{Background}
\label{sec:background}

In this section, we provide the background information on Few-shot learning and Siamese neural networks. In App.~\ref{app:nn} we provide further background information on neural networks.

\subsection{Few-shot Learning}

\begin{figure}[tb]
	\centering
	\includegraphics[width=\columnwidth,trim={4cm 0 4cm 0},clip]{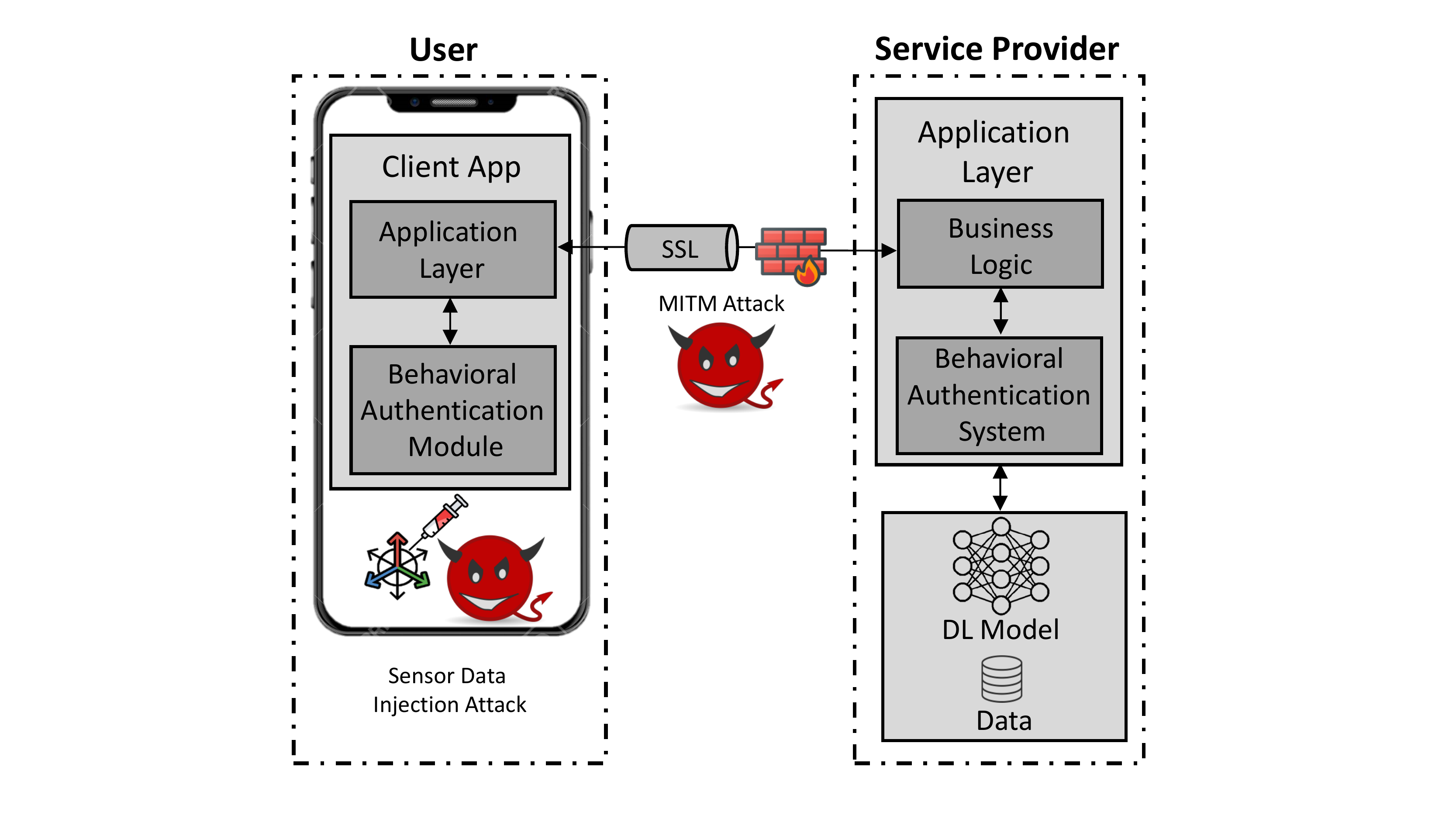}
	\caption{Threat Model}
	\label{fig:threst}
\end{figure}
 
\begin{figure*}[htp]
	\centering
	\includegraphics[width=0.95\textwidth]{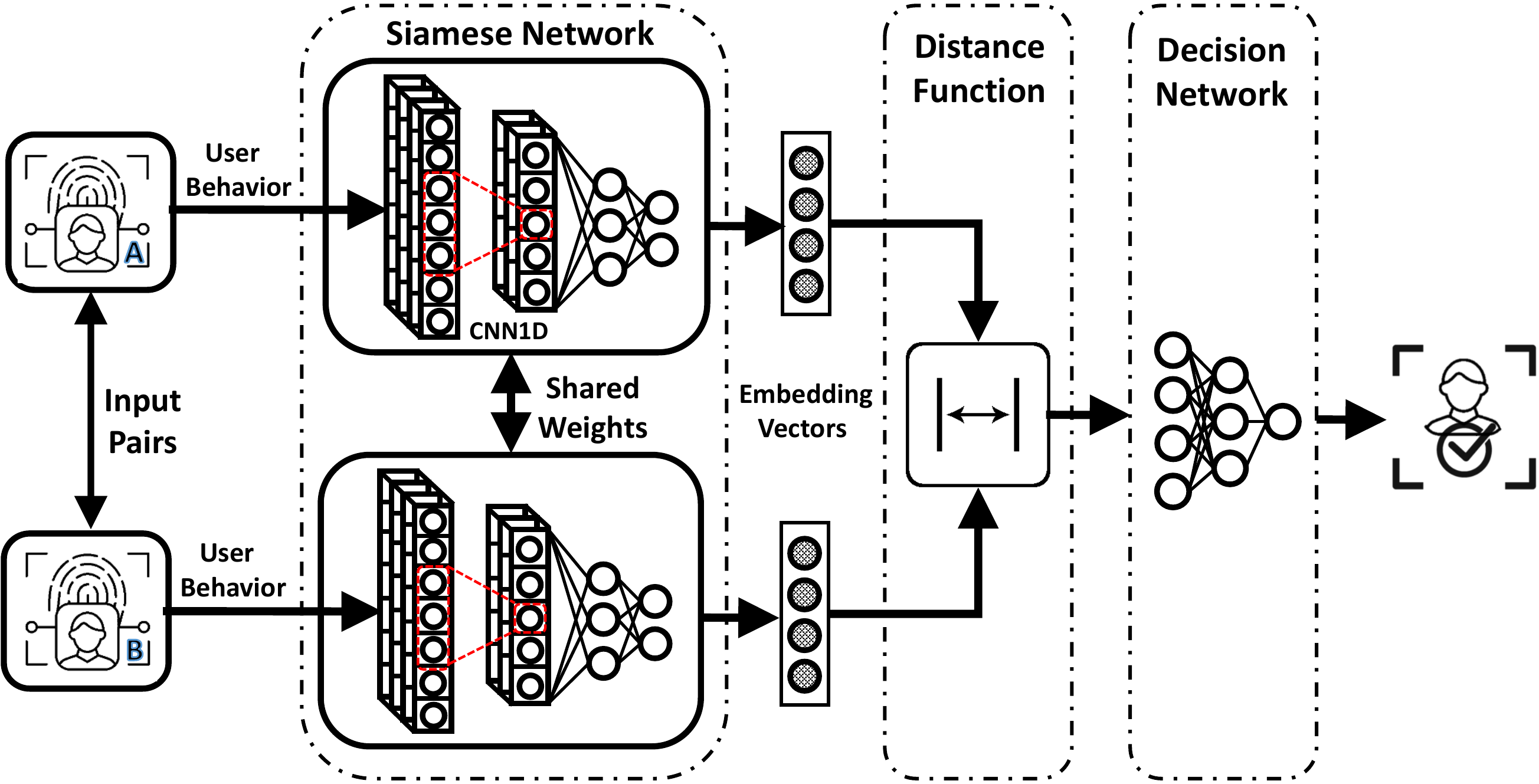}
	\caption{ An abstract view of \ourname at inference time}
	\label{fig:highlevel}
\end{figure*}
Few-shot learning is a machine learning method where models are trained with only a limited number of samples for each class. The common practice for machine learning applications is to feed as much data as the model can take. This is because feeding more data to most machine learning (especially deep learning) applications enables the model to generalize better. However, few-shot learning aims to build precise models with less training data. The fundamental idea is to learn new concepts and generalize tasks from only a few examples. 
One approach to few-shot learning includes Siamese networks which involve learning an embedding space to compare classes. Few-shot learning has gained traction in areas such as computer vision, natural language processing, audio processing, robotics and medical applications where a large number of classes exist and labeled data is scarce \cite{10.1145/3386252}.

\subsection{Siamese Networks}
Siamese Networks have been utilized for few-shot classification or representation learning problems in which only a limited number of samples are available for each class. They can also be applied to classification problems that suffer from lack of enough data where only one sample per class is available (one-shot learning). A Siamese Network is a tandem of two identical of any given network architectures (i.e., MLP, RNN, or CNN) with the objective of finding the similarity or a relationship between two comparable observations. Both sub-networks share the same parameters (weights $w$ and biases $b$). The objective of the training is to extract similar features (i.e., embedding vectors) if the two input samples belong to the same class, while extracting differing features if the two input data belong to the different classes where the similarity between vectors is typically quantified by a \textit{p-norm} distance metric, with the Euclidean distance being the most commonly used one.

\section{System and Threat Model}
\label{sec:threat}

Our system model, as shown in Figure~\ref{fig:threst},  includes users who access online services through their mobile apps and a service provider after an authentication process. In this section, we describe the assumptions we make regarding the attacker capabilities and the threat model for our behavioral authentication system. 
The main objective of the adversary is to bypass the authentication system and impersonate the end-user while authenticating into a sensitive service (e.g., mobile banking). In particular, we make the following assumptions:

\begin{itemize}
    
    \item    The authentication model is trained and maintained in a data center owned by the service provider. 
    In typical user authentication applications, to complete a successful and secure authentication process, the service provider is trusted. Since the authentication model is trained at the service provider, this will avoid adversarial machine learning such as model poisoning, or privacy attacks.

    \item The adversary can compromise the user’s mobile device, e.g., by installing malware or exploiting vulnerabilities. 
    \item     Aligned with existing work [66], we assume that sensor data is trustworthy as its integrity can be protected by hardware security technologies such as ARM TrustZone [52]. The Behavioral Authentication Module (BAM) in the client application, Fig 1., can also run in the secure world of an ARM TrustZone. The motion sensors can be exclusively assigned to the secure world (and the BAM) to avoid any access to sensor readings from outside of the secure world. 
    
    In case the OS is compromised, the adversary can still not access the sensor, since reading from the sensor is only possible from the secure world.

    \item    Side-channel attacks to extract user behavioural data from device sensors are orthogonal to this work and can be mitigated with sensor data obfuscation techniques~\cite{Das2016, Shrestha2016}.
    \item    The communication channels are secured through the use of standard secure communication protocols such as SSL/TLS, and hence are assumed not to be affected by the adversary. Also, the adversary cannot tamper with the data that the BAM already reads from the sensor and they can also not conduct MITM or replay attacks on the device. To perform such attacks the attacker needs to sniff the secure communication channel, which requires breaking, e.g., SSL/TLS channel.
\end{itemize}

To authenticate a new user, AuthentiSense collects a number of enrollment samples (e.g., 3). After the initial setup, AuthentiSense compares the (current) user samples against the enrollment samples. If the samples match (collected samples in the initial setup and current sample belong to the same user), AuthentiSense produces similar embedding vectors, therefore, the Euclidean distance between the computed embedding vectors in the latent space is minimized. AuthentiSense performs a binary classification based on the distance between the embedding vectors. As result of the classification, it outputs the binary label “1” leading to successful user authentication.

AuthentiSense authenticates users with high reliability (see Section V-D); but, in the event of continuous authentication failure, after three unsuccessful attempts, the AuthentiSense automatically falls back to a passive authentication technique (i.e., password) to enforce security.
\section{\ourname}
\label{sec:design}

In this section, we first outline the high-level architecture of \ourname and then describe its components in more detail.

\subsection{High-level Overview} \label{sec:high-level}

The high level architecture of our system is depicted in Figure~\ref{fig:highlevel}. The system architecture involves the following components: The Siamese networks (CNN-based), distance function and decision network. At a high-level, the Siamese network functions as a feature extractor generating embedding vectors from input data. Then a Euclidean distance between embedding vectors are computed by the distance function and fed to the decision network for classification. 

\noindent \textbf{Siamese network.} It consists of two identical sub-networks having the same structure and sharing weights. Each sub-network processes one user behavior and function as a feature extractor learning a meaningful representation of input samples. Two recorded user behavior samples are fed to the sub-networks and transformed into the embedded space learned by the Siamese network.

\noindent \textbf{Distance function.} It computes the Euclidean distance between the computed embedding vectors in the latent space. The Euclidean distance here can also be seen as the L$_2$-norm of the distance vector. The objective of the Siamese network is to maximize the calculated value of this function for negative pairs, i.e., when the input samples belong to different users, and minimize the distance value for positive pairs, i.e., samples from the same user.

\noindent \textbf{Decision network.} It consists of fully-connected layers which performs a binary classification based on the distance between the embedding vectors in the latent space. As result of the binary classification, the decision network outputs a binary label, “1” if the captured behavior samples belong to the same user, and “0” otherwise.

\subsection{Design}
\label{sec:detailed_design}

AuthentiSense is mainly designed for real-world deployment where our focus was on a mobile banking application. We stress that to train the AuthentiSense, diverse user behaviors (i.e., different genders, ages, and occupations) were captured on a real mobile application from a bank where the users provided their signed consent before data collection. Moreover, to demonstrate the practicality of AuthentiSense, the most frequently used functions in the mobile application (according to an analysis conducted on the usage patterns of the bank customers [40]) were identified and used to simulate the user behavior. \ourname consists of three components, each of which has a precise task. In the following, we elaborate on each component in more detail. 

\begin{figure}[t]
	\centering
	\includegraphics[width=0.46\textwidth]{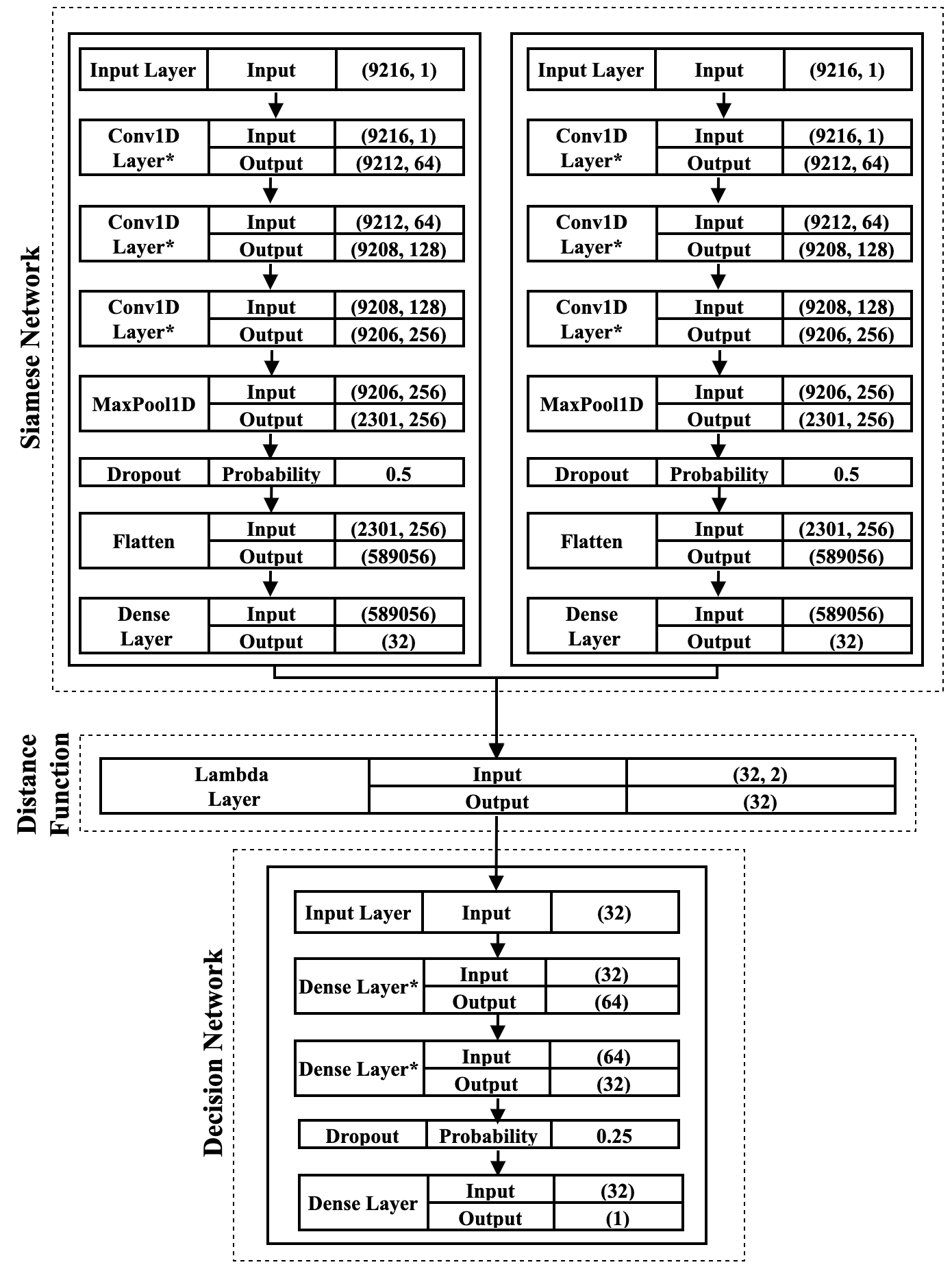}
	\caption{Model Architecture: Siamese Neural Network followed by the Distance Function and the Decision Network. The layers with $*$ are followed by a Batch Normalization layer.}
	\label{fig:arch}
\end{figure}

\noindent \textbf{Siamese Network.} We utilize Siamese neural network to extract highly discriminative features for input data that can distinguish the behavior between genuine and impostor users. Particularly, the Siamese network aims to learn information-rich transformation of the input data into an embedding (latent) space that can preserve distance relation between input data. Suppose we are given a pair of recorded behavior samples as input data; the objective is to map them to an embedding space where the embeddings of two input samples from the same user are closer together and two input data from different users are far apart.  The Siamese network architecture, as depicted in Figure~\ref{fig:arch}, is made of two identical sub-networks (CNN1D, cf. App.~\ref{app:cnn1d}) which share weights and biases. Each sub-network processes one of the input behavior samples and works as a feature extractor and outputs an encoding of recorded behavior (embedding) using the same filters as the other. We perform L$_2$-normalization to normalize the embedding vectors and map them to the surface of n-dimensional hyper-sphere of radius 1. This allows to compare the similarity between different inputs by distance between two embedding vectors. As all the embeddings will reside on the surface, this prevents that embeddings with a high L$_2$-norm distract the decision network.

\noindent \textbf{Distance function.} 
In order to train the Siamese network, we define a triplet loss function (cf. \sect\ref{sec:impl}) which, for a given pair of input data, takes the distance between the two output embedding vectors from the two sub-networks and regulates large or small distances.
The generated embedding vectors are subjected to a distance function (created using a Lambda layer~\cite{lambda}) which computes the L$_2$-distance.
Basically, this distance between embedding vectors should be large enough when the input samples belong to different users but small when they belong to the same user.
Among several distance functions, we opted for the L$_2$-distance. 
We evaluated other distance metrics (i.e., L$_1$-distance, and cosine distance), but L$_2$-distance showed the best results.
Utilizing the L$_2$-distance, we fine-tuned the network parameters (i.e., weights) using back propagation~\cite{rumelhart1986learning}. 

\noindent \textbf{Decision Network.} A fully-connected decision network followed by the distance function makes the classification decision based on the distance between the embedding vectors in the embedding space.  The role of the decision network is to solve a binary classification problem and it outputs “1” if the behavior samples belong to the same user, and “0” otherwise. We design a feed forward neural network with three subsequent dense layers with decreasing width in terms of neurons, and ReLU activation function along with a L$_2$-norm as kernel regulizer. The first-two layers are then followed by a batch normalization layer for regularizing the activations of the prior layer, so that they have mean close to 0 and standard deviation close to 1. The last dense layer, followed by a dropout layer with probability of 0.25, is made of a single unit with sigmoid activation function for being consistent with a Bernoulli output distribution. 

As an input, the decision network receives the distance vector of two embedding vectors and produces in output a probability value (i.e., in range $[0,1]$), which indicates how likely the input data belongs to the same user. 
Since the point of Equal Error Rare (EER) represents the best classification threshold (because it is the point where False Acceptance and False Rejection rates meet, cf. \sect\ref{sub:metrics}), we set the threshold for classification to the point of EER for a set of validation data that were not used for training the neural networks (cf. Fig.~\ref{fig:eval-eerComparison:training}), to decide if we will consider the prediction of the decision network as correct.

\subsection{Implementation}
\label{sec:impl}
In the following, we explain sample preparation to train the Siamese and decision networks. We then elaborate on model building and describe our training strategy and model hyperparameter search and tuning in more detail.

\noindent \textbf{Sample preparation.} The Siamese network learning process is based on comparing pairs of behavior samples, namely positive and negative pairs. In positive pairs, two samples belong to the same user, while in negative pairs two samples are from different users. During the model training phase multiple samples of these pairs are fed to the model to minimize the L$_2$-distance between the embeddings of positive pairs, and to maximize the L$_2$-distance for negative pairs. One way of selecting the positive and negative pairs is a random selection. This na\"{i}ve approach has been shown to be  good enough for the positive pair selection.
However, for negative pairs this na\"{i}ve approach is problematic as many samples from different users differ significantly, while it is more efficient for training to focus on negative pairs that the network fails to distinguish. 
Sophisticated techniques such as triplet loss can be utilized for sample preparation, in which three samples simultaneously are used to optimize every training step. In triplet sampling, the first two samples are positive and the last one corresponds to a negative sample. The goal is to minimize the L$_2$-distance in the embedding space between positive samples and maximize the L$_2$-distance between positive and negative samples. 

\begin{figure*}
	\centering
	\includegraphics[width=0.81\textwidth]{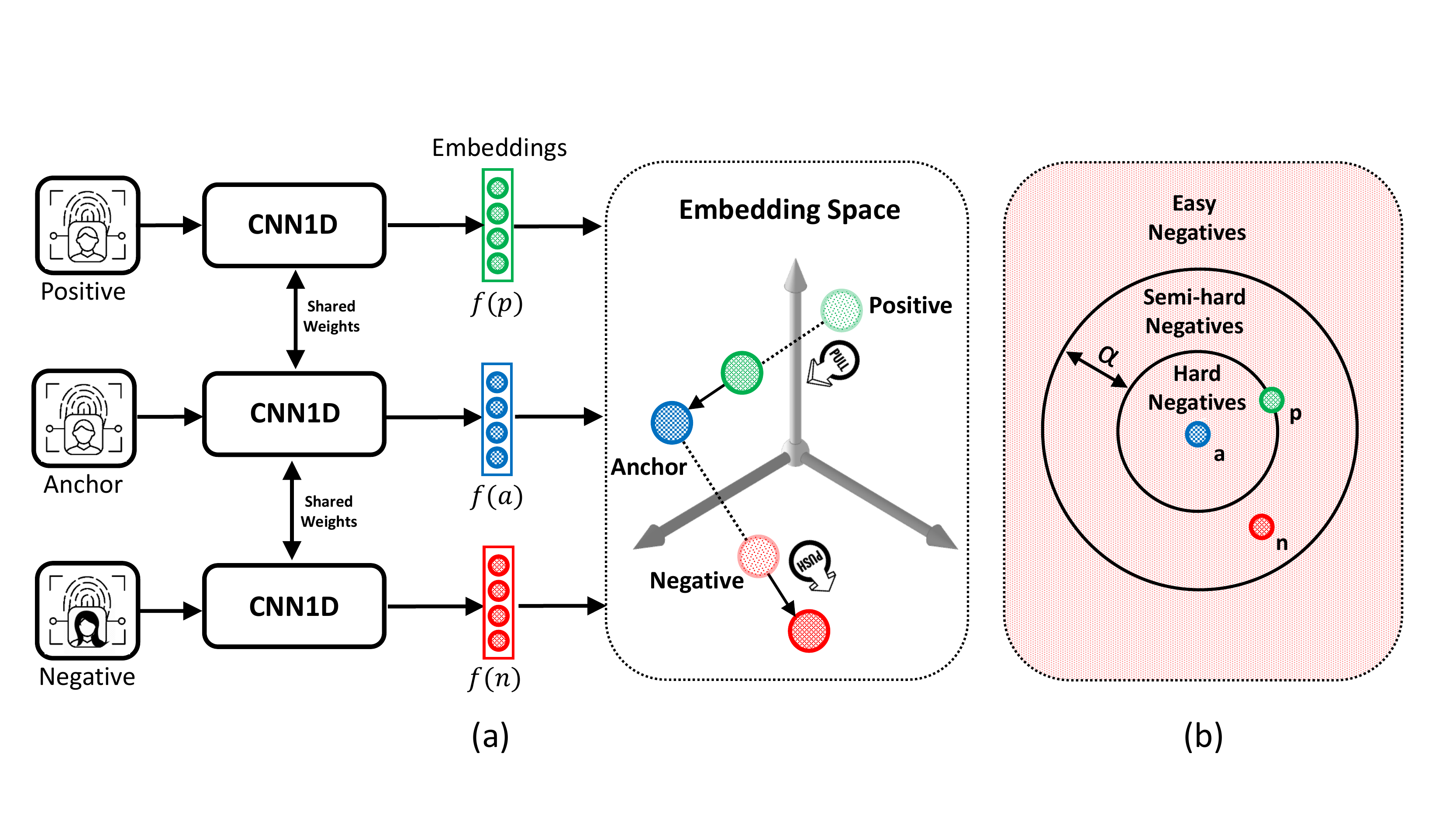}
	\caption{Training using triplet loss: figure (a) demonstrates how triplet loss pushes positive and anchor samples toward each other and pulls negative sample from anchor. Figure (b) shows Semi-hard triplet selection scheme.}
	\label{triplet_training}
\end{figure*}

\noindent \textbf{Model building.} The deep learning part of \ourname uses a Siamese Neural Network (NN) and a decision network. The Siamese NN consists of 2 sub-networks, arranged in the Siamese architecture by assembling layers (in this case, Conv1D, Max Pooling1D, Flatten and  Dense layers). Afterwards, the outputs of the two sub-networks are piped through our custom distance function (using a Lambda layer), as shown in Figure~\ref{fig:arch}. In the end, we append the decision network to the Siamese network and build our classifier by stacking fully connected Dense layers.

We use three subsequent convolutional layers with an increasing number of filters (from 64 to 256) in which data is expanded in depth, and with a kernel size that is reduced in the last layer (from 5 to 3) for a more fine-grained computation. Since we are dealing with non-normalized data, after each convolutional layer we add a batch normalization layer to normalize the inputs to a layer for each mini-batch of data. This has the effect of stabilizing the learning process and dramatically reducing the number of training epochs required to train the Siamese network~\cite{ioffe2015batch}. Furthermore, each convolutional layer includes a kernel regularizer (L$_2$-norm) with a value of $10^{-3}$.

After the convolutional layers, we use a 1-dimensional Max Pooling layer for down-sampling the input representation by taking the maximum value over a spatial window of a specified size (in our case 4). Then, we flatten the output of the convolutional block and feed it to the last fully connected (i.e., dense) layer with no activation which is responsible for producing the embedding representation of the input as an array of fixed size (length of 32×1). 

\noindent \textbf{Training Strategy.} The training procedure depends on the sample generation strategy (i.e., pairwise or triplet). For the pairwise training a contrastive~\cite{1640964}
loss is used (cf. App.~\ref{sec:appendix}) while triplet loss~\cite{tf_addons} is utilized for triplet training.
Triplet loss learns both positive and negative distances simultaneously

and focuses more on negative samples that are hard to distinguish reducing the number of easy-distinguishable negative samples. This helps to reduce the risk of overfitting compared to pairwise training. We opt for triplet training, therefore, we train our Siamese network with triplets loss. The objective of triplet loss is to ensure that two samples with the same label have their embeddings close together in the embedding space, while embeddings of two samples with different labels are far away from each other.

To train the network, we sample triplets of desired batch size. As the name implies, three input samples are needed, which are called: i) \textit{Anchor} which is a sample input data, ii) \textit{Positive} that is just another variation of the anchor, 
and iii) \textit{Negative} which is a different sample from above two similar samples. This helps our model learn dissimilarities with the anchor sample. 
Positive and negative samples are passed individually to the Siamese network, as triplets are mined online~\cite{schroff2015facenet}. As depicted in Fig.~\ref{triplet_training} the network learns to decrease the distance between the anchor and positive, while increase the distance between anchor and negative such that the difference of the two distances would reach to Alpha $\alpha$ which is a pre-defined hyper-parameter (Eq.~\ref{eq:trip}). 

The alpha parameter in Eq. 1 aims to ensure that the difference between the anchor-positive distance (d(a,p)) and the anchor-negative distance (d(a,n)) is at least as big as a margin equal to alpha to discourage the model from collapsing to trivial solutions where f(a) = f(p) = f(n) which would satisfy Eq. 1. The triplet loss is defined as follows:

\begin{equation} 
    \mathcal{L}(a,p,n) = max(||f(a)-f(p)||^2-||f(a)-f(n)||^2 +\alpha,0)
     \label{eq:trip}
\end{equation}

In the triplet loss function, the term $||f(a)-f(p)||$ is the distance between the anchor and positive and the term $||f(a)-f(n)||$ is the distance between the anchor and negative. The value of the first term is learned to be smaller while the second term to be bigger. If their subtraction is smaller than minus alpha, the loss would become zero and the network parameters would not be updated at all. During the training, we minimize this loss, which pushes $||f(a) - f(p)||$ to 0 and $||f(a) - f(n)||$ to be greater than  $||f(a) - f(p)||$ + $\alpha$.
Based on the definition of the triplet loss, there are three categories of triplets:
\begin{itemize}
\item Easy triplets which have a loss of 0, because  $||f(a) - f(p)|| + \alpha \le ||f(a) - f(n)||$
\item Hard triplets where the negative is closer to the anchor than the positive, i.e., $||f(a) - f(n)|| \le ||f(a) - f(p)||$
\item  Semi-hard triplets where the negative is not closer to the anchor than the positive, but which still have positive loss: $||f(a) - f(p)|| \le ||f(a) - f(n)|| \le ||f(a) - f(p)|| + \alpha$
\end{itemize}

Each of these triplet definitions depends on where the negative sample is located, relatively to the anchor and positive. The categorization (Easy, Hard, and Semi-hard triplets) can therefore be applied to the negative pairs analogously: Hard negatives, Semi-hard negatives or Easy negatives, as shown in Fig.~\ref{triplet_training}.
So theoretically, in order to ensure the best effect of network training, we need to choose Hard triplets or Semi-hard triplets. The Hard negatives deliver better losses for model optimization and lead to a strong convergence, but in practice they might be too aggressive and collapse the loss function. Therefore, we opt to use a semi-hard mining strategy in model training. 

After defining the loss on triplets of embeddings and observing that some triplets are more useful than others, meaning their loss values help network for a better weight optimization, we select online triplet mining~\cite{schroff2015facenet} approach to mine triplets. Unlike offline triplet mining where the data is fed as a triplet-form input to the network, in online triplet mining triplets are computed during training within each batch of data. The idea behind online mining is to dynamically compute useful triplets on the fly, for each batch of data. Given a batch of $B$ samples, it computes the $B$ embeddings and then can find a maximum of $B^3$ triplets\footnote{However, since each triplet must consist of an anchor, a sample from the same user of the anchor (positive pair), and a sample from a different user (negative pair), many of these na\"ive combinations are invalid, i.e., do not contain a negative and a positive pair, s.t., the actual number of valid triplets is smaller than $B^3$ and depends on the batch.} and selects the triplets for training that are semi-hard.

As already described, we use semi-hard triplet mining to train the network, as semi-hard triplets are not as difficult as hard triplets to learn, but still provide useful information. The mining strategy that we adopt computes triplets online from a batch of randomly drawn samples, therefore, it is impossible to pre-determine the number of triplets.

We also stress that Figure 2 represents an abstract view of the system at inference time, while Figure 4 illustrates the AuthentiSense at training time. We utilized online triplet mining [60] to train our network and unfolded the network structure, in online mining, to explain the inner workings of our system.

\noindent \textbf{End to end model parameters optimization.} Having trained the Siamese network with triplet loss, we perform an end to end parameter (i.e., weights) optimization  of the decision network using standard backpropagation on the predictions of the entire network including both the Siamese and decision networks. This includes the following steps: First, the weights of the Siamese network are frozen (it is used as a feature extractor),
then fully connected decision network is appended to the Siamese network 
and trained using a binary cross-entropy loss function to optimize the network's weights.

\begin{table}[tb]
\centering
\begin{tabular}{l|c}
\textbf{Variable} & \textbf{Setting} \\ \hline
Optimizer & Adam, SGD \\
Learning Rate & $0.1, 0.05, 0.01, 0.005, 0.001, 0.0005$ \\
Batch Size & 64, 128, 256, 1024 \\
Margin (Alpha) & $1, 0.5, 0.3, 0.1, 0.05, 0.03, 0.01$
\end{tabular}
\caption{Hyperparameter search for the Siamese Network}
\label{tab:param_search_snn}
\end{table}

\begin{table}[tb]
    \centering
    \begin{tabular}{l|c}
   \textbf{Variable} & \textbf{Setting} \\ \hline
    Optimizer & \multicolumn{1}{c}{Adam, SGD}  \\
    Learning Rate & $0.1, 0.05, 0.01, 0.005$  \\
    Batch Size & 32, 64, 128  \\
    \end{tabular}
    \caption{Hyperparameter search for the Decision Network}
    \label{tab:param_search_decnet}
\end{table}

\noindent \textbf{Hyperparameter tuning and network configuration.} After constructing the model architecture, we take a Grid Search \cite{scikit-learn} approach and loop through pre-defined hyperparameters, as shown in Tables \ref{tab:param_search_snn}, and \ref{tab:param_search_decnet}, to choose a set of optimal hyperparameters for the learning algorithms that maximize the model performance. In particular, we investigate multiple options concerning the choice of the optimizer, learning rate, batch size, and the margin value (Alpha) for the Semi-Hard Triplet Loss function.
At the end, we select the best performing parameters that maximize the performance of the models, as shown in Tables \ref{tab:SNN_hyperparms} and \ref{tab:DecisionNet_hyperparms} \mbox{and use them for training our networks.}
\section{Evaluation}
\label{sec:setup}
In the following we describe the evaluation and also show different alternatives for the design of \ourname .
\subsection{Dataset}\label{subsec:dataset}

To conduct our experiments, we use the DAKOTA Dataset~\cite{incel2021dakota}, more specifically, the recorded motion sensor values (i.e. accelerometer, gyroscope and magnetometer) for 45 users while using a mobile banking smartphone app. Out of the 45 users, we randomly selected 35 users for training, 3 users as validation data to determine the classification threshold, and 7 users for testing. Data was recorded for every user in 5 sessions for each of the postures \emph{sitting}, \emph{standing} and \emph{phone on table}. Therefore, each user in the training set has 15 sessions, each 90 seconds in length. So it takes only a few volunteer users to train the feature extractor. Furthermore, we have mainly focused on having a few-shot solution for new users joining after the system is deployed. We could have asked the volunteers to provide as much data as we wanted. However, for real-world users, we can only expect to collect a few samples (Table V). As the raw data was sampled with a non-constant frequency and non-uniform starting and ending timestamps for each sensor, we resampled the sensor data for every session at a rate of 5 ms,
taking the mean in areas where the data was downsampled and linearly interpolating in areas where the data was upsampled. To obtain a pool of samples, we then ran a sliding window of fixed length (1, 3, 5, 10 and 15 seconds) with step size equal to 1/10th of its length over the data of every recorded session and labeled it with the corresponding user. For each window, the values of every axis of each motion sensor are concatenated to shape a one-dimensional array 
that is later used to construct positive and negative examples. The pool of windows was then shuffled and split up into batches for the training and testing sets to be used for the Siamese network. For the decision network, an equal number of window pairs by the same user (positive example) and different users (negative example) was randomly sampled to generate 50,000 and 10,000 pairs which were then batched for training and testing, respectively.
\begin{table}[tb]
\centering
\begin{tabular}{l|c}
\multicolumn{1}{c|}{\textbf{Hyperparameter}} & \multicolumn{1}{c}{\textbf{Setting}} \\ \hline
Optimizer & Adam \\
Learning Rate & $10^{-3}$ \\
Batch Size & $64,128,256$ \\
Loss Function & Semi-Hard Triplet Loss \\
Margin (Alpha) & 0.03 \\
\end{tabular}
\caption{Siamese Network hyperparameter settings}
\label{tab:SNN_hyperparms}
\end{table}

\begin{table}[tb]
\centering
\begin{tabular}{l|c}
\multicolumn{1}{c|}{\textbf{Hyperparameter}} & \multicolumn{1}{c}{\textbf{Setting}} \\ \hline
Optimizer & Adam \\
Learning Rate & $10^{-4}$ \\
Batch Size & 64 \\
Loss Function & Binary Cross-entropy \\
\end{tabular}
\caption{Decision Network hyperparameter settings}
\label{tab:DecisionNet_hyperparms}

\end{table}

\subsection{Evaluation Metrics}
\label{sub:metrics}
The performance evaluation of \ourname is performed using common metrics. Each metric is based on the number of correctly (TP) 
and incorrectly (FN) classified benign authentication events as well as the number of correctly detected (TN)  and unrecognized (FP) attack attempts. We use the following metrics to evaluate the effectiveness of \ourname.

\noindent
\textbf{False Acceptance Rate (FAR)} represents the risk to accept attack
attempts and is defined as follows:

\begin{equation}
\text{FAR} = \frac{ \text{FP} }{ \text{FP} + \text{TN} }
\end{equation}
\noindent
\textbf{False Rejection Rate (FRR)} analogously represents the risk to
mistakenly decline benign attempts. It is defined as:

\begin{equation}
\text{FRR} = \frac{ \text{FN} }{ \text{FN} + \text{TP} }
\end{equation}

\noindent
\textbf{\fscore}: The \fscore is harmonic mean of Precision and Recall. The Precision calculates the ratio between the number of positive samples classified correctly, and the total number of samples classified as positive (Eq.~\ref{pr}). Recall defines how many positive samples have been identified correctly overall (Eq.~\ref{re}). 

\begin{equation} \label{pr}
    \text{Pr}=\frac{\text{TP}}{\text{TP}+\text{FP}}
\end{equation}

\begin{equation} \label{re}
    \text{Re}=\frac{\text{TP}}{\text{TP}+\text{FN}}
\end{equation}
The \fscore is then defined as:
\begin{equation}
\text{\fscore} = 2\cdot \frac{ \text{Pr} \cdot \text{Re} }{ \text{Pr} + \text{Re} }
\label{eq:f1score}
\end{equation}

\noindent
\textbf{Area Under ROC Curve (AUC)}: The above-mentioned metrics depend on a threshold that converts the predicted probabilities into binary predictions (accept and reject). The Receiver Operating Characteristic (ROC) curve plots the model's True Positive Rate $\text{TPR}=1-\text{FRR}$
against FAR for different threshold values, as illustrated in Fig.~\ref{fig:roc}. The resulting area between such curve and the X-axis at an interval of $[0,1]$ is termed AUC and expresses the model's capability of correct classification. 

\noindent
\textbf{Equal Error Rate (EER):} The EER is the value of FAR (and equally FRR) at a
threshold value where both FAR and FRR embody equal values.

\subsection{Experimental Setup}
\label{sub:flsetup}
All the experiments were conducted on a server running Debian 10, with 1 TB of memory, 64 physical cores/128 threads, provided by an AMD EPYC 7742 processor, and 4 NVIDIA Quadro RTX 8000. We leveraged TensorFlow 2.4.0~\cite{tensorflow} to implement the neural networks. We trained our Siamese network using Semi-Hard triplet loss (cf.~\sect\ref{sec:design}). We used  Contrastive loss as a baseline to compare the results of this loss function with results obtained from the triplet training in \ourname. For all the loss functions, we used the implementation provided by the Tensorflow Addons (TFA) library~\cite{tf_addons}.

\subsection{Evaluation Results}
\label{sec:results}
To evaluate the effectiveness of \ourname, we first discuss the performance of \ourname, 
and then we compare it against the baseline to see how the choice of loss function for training the Siamese network can influence the performance. We also conduct a performance comparison with individual sensor modalities and fusion of modalities. 

\subsubsection{Performance of \ourname}
To verify the capabilities and performance of \ourname, we conduct a number of experiments with different setups in the training and recognition phases. We perform a systematic evaluation with different variables such as authentication window time (interaction time required for authentication purpose) and the number of known user samples (from previous history) to compare with (n-shot).

As shown in Tab.~\ref{tab:user-based-auth:tripplet}, \ourname achieves a \fscore of 0.97 in 3-shot verification just in \textit{1 second} of interaction with the user, meaning that for user authentication it only needs 3 enrollment samples.
\ourname can also verify users in \textit{1-shot} after 1 second of user interaction at the cost of 0.02 drop in the model performance (\fscore=0.95). The table also shows that the results vary slightly for an authentication window length of 1s for 4- and 5-shot verification (\fscore=0.96). 

Moreover, Tab.~\ref{tab:user-based-auth:tripplet} shows that for 5-shot verification, \ourname effectively authenticates the users for all authentication window lengths. This demonstrates the advantage of the few-shot learning approach that allows \ourname to obtain a satisfactory description of user behavior from the collected information (i.e., the motion patterns), resulting in reliable authentication results, even for corner cases with small number of (n) shots and short interaction time.

Especially, the shorter authentication window time allows \ourname to perform verification very quickly at recognition stage with only limited amount of data collected during user interaction (t=1s).
This confirms the usability of \ourname in practice. The users are not required to interact for a long time with their devices to collect enrollment samples needed for authentication process.

\begin{table}[tb]
\centering

        \centering
        \begin{tabular}{ll|l|l|l|l|c|}
        \cline{3-7}
                                                          &    & \multicolumn{5}{c|}{\authentWindowLength (Sec.)}                                                                    \\ \cline{3-7} 
                                                          &    & \multicolumn{1}{c|}{1} & \multicolumn{1}{c|}{3} & \multicolumn{1}{c|}{5} & \multicolumn{1}{c|}{10} & 15 \\ \hline
        \multicolumn{1}{|l|}{\multirow{5}{*}{\STAB{\rotatebox[origin=c]{90}{\nComparison}}}} & 1 & 0.95 & 0.88 & 0.91 &0.85 & 0.85\\ \cline{2-7}
        \multicolumn{1}{|l|}{} & 2 & 0.96 & 0.90 & 0.92 &0.90 & 0.88\\ \cline{2-7}
        \multicolumn{1}{|l|}{} & 3 & 0.97 & 0.91 & 0.94 &0.92 & 0.82\\ \cline{2-7}
        \multicolumn{1}{|l|}{} & 4 & 0.96 & 0.92 & 0.92 &0.94 & 0.95\\ \cline{2-7}
        \multicolumn{1}{|l|}{} & 5 & 0.96 & 0.93 & 0.94 &0.94 & 0.95\\ \hline
        \end{tabular}
    \caption{\fscore for triplet training on test set.}
    \label{tab:user-based-auth:tripplet}
\end{table}

\begin{figure}[tb]
  \centering
 \subfloat[\label{fig:eval-eerComparison:training}Calculation of the Equal Error Rate (EER).]{
      \includegraphics[width=.5\textwidth]{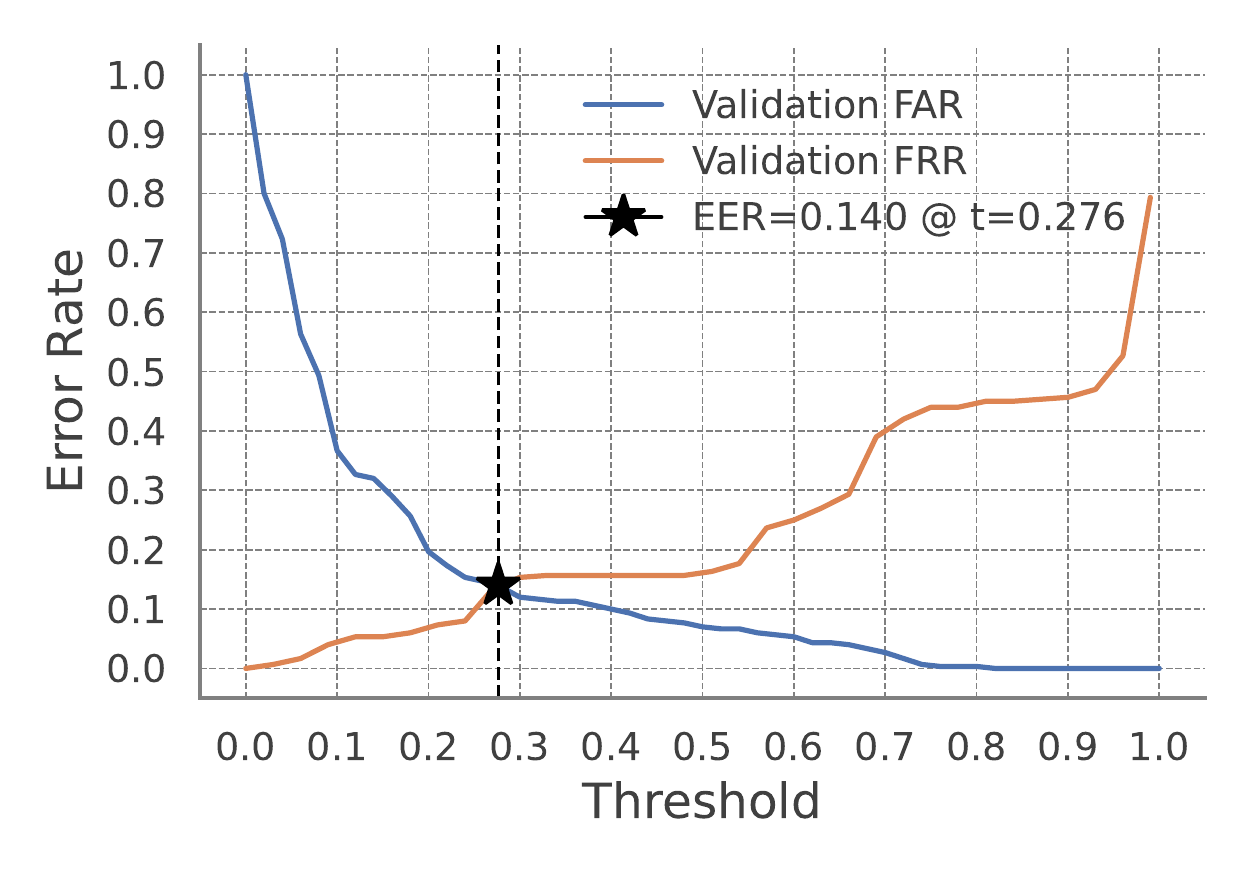}}
  
  \vspace{2em}
   \subfloat[\label{fig:eval-eerComparison:test}Calculation of FAR and FRR.]{
      \includegraphics[width=.5\textwidth]{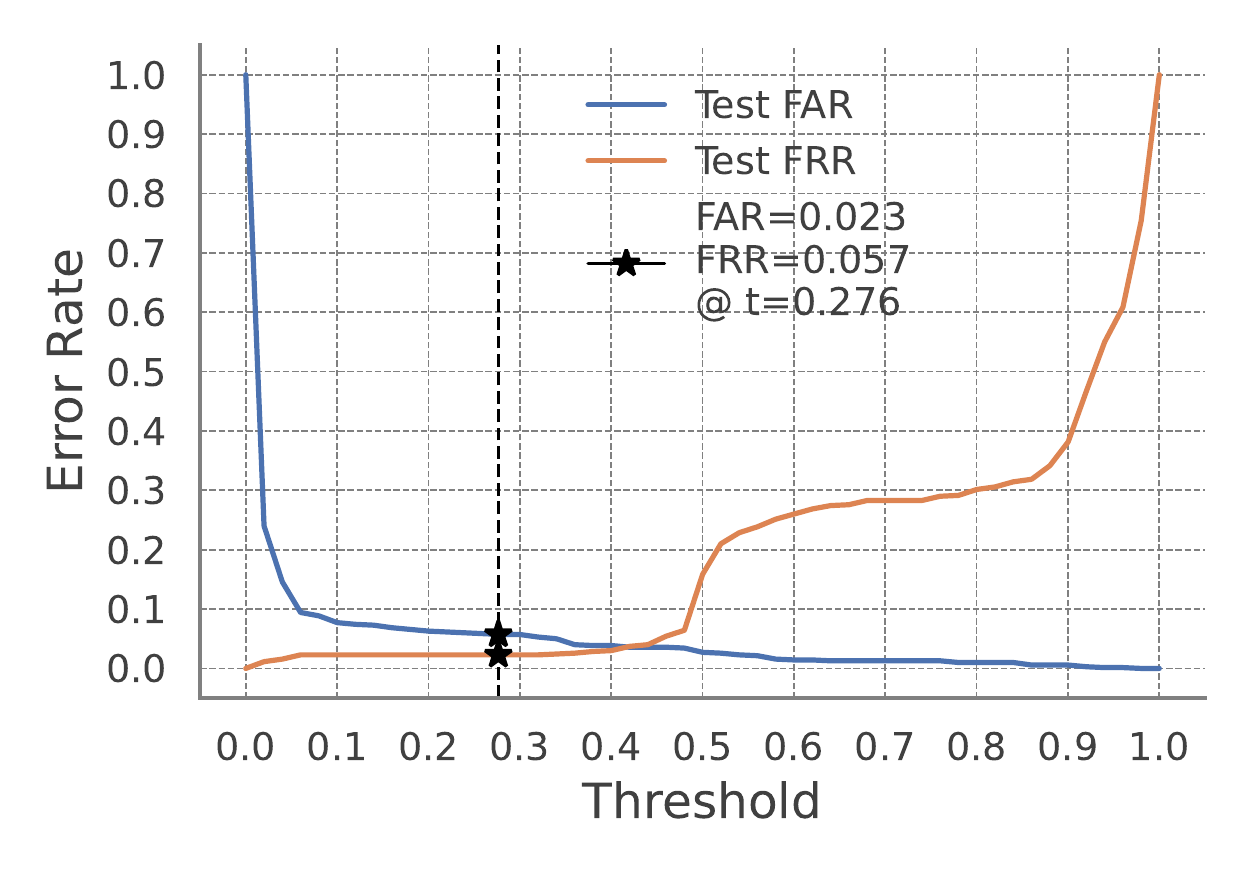}}
  \vspace{0.9em}
    \caption{\label{performance} Calculation of the  EER=0.140 and threshold (t = 0.276) on validation data (a). Computation of FAR=0.023 and FRR=0.057 at the threshold point (t=0.276) on testing data (b).}
\end{figure}

Figure~\ref{fig:eval-eerComparison:test} depicts the FAR and FRR for different classification thresholds. 
For \ourname, the point of EER for the validation data, as shown in Fig.~\ref{fig:eval-eerComparison:training}, represents the best threshold (t=0.276) to choose, because it is the point where FAR and FRR are equal. Having applied this threshold on the unseen test data, \ourname achieves a FAR=0.023 and an FRR=0.057. It should be noted that FAR is significantly lower than FRR. Since a false-reject results in an additional request for manual authentication while a false-accept leads to an unauthorized access, thus, having  as low FAR as possible is more important for \ourname. 

\begin{figure} [tb]
	\centering
  \includegraphics[width=.45\textwidth]{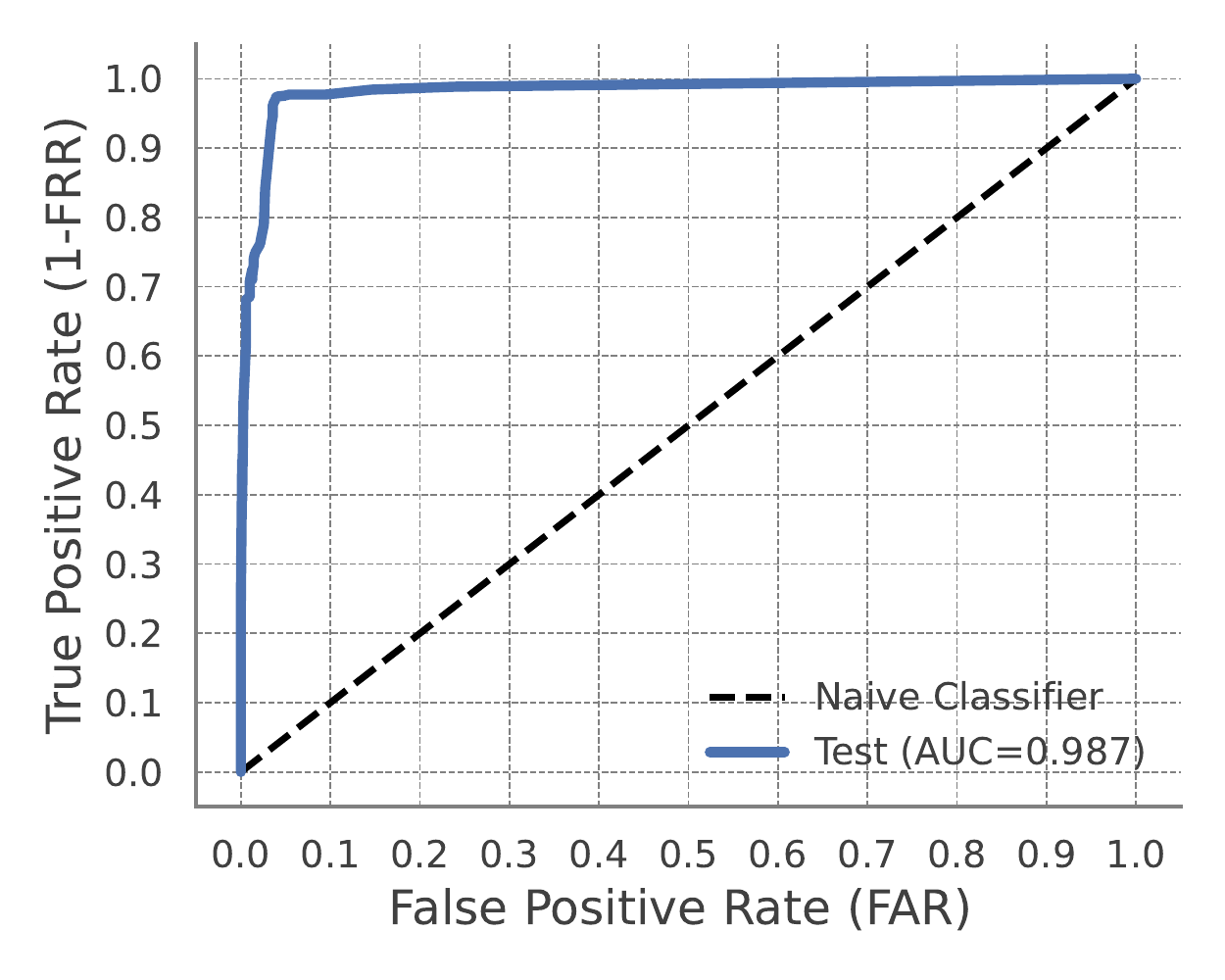}
	\caption{ROC curve: The area under curve for test samples which serves as a measurement for the model performance.}
    \label{fig:roc}
\end{figure}

Moreover, Fig.~\ref{fig:roc} illustrates the Receiver Operating Characteristic (ROC) curve for test samples. The ROC curve shows the dependency between the FAR, FRR and the system’s detection threshold. The Area Under Curve (AUC), ranges from 0.5 (random guessing) to 1 (perfect classification), aggregates the system’s performance at all threshold settings and acts as an indicator of the model performance. As it can be seen in the figure, \ourname obtains AUC=0.987 for unseen test data.

\subsubsection{Comparison against baseline}
\label{sec:ablationstudy}
We perform an experiment where we set the contrastive loss as a baseline to compare the model performance in terms of \fscore against triplet training. 
Table~\ref{tab:user-based-auth} shows the performance of \ourname for contrastive loss function while training its Siamese Network. As shown in the table, the triplet loss outperforms contrastive loss function, for almost all numbers of shots (\nComparison) and all authentication window times. Also the best \fscore for the triplet loss (0.97), as shown in Tab.~\ref{tab:user-based-auth}, is higher than the best value of the contrastive loss (0.93). The better efficiency becomes especially visible for challenging authentication scenarios.
For example, for an interaction time of 1s and \nComparison= 3, the contrastive loss baseline achieves an \fscore of only 0.92, while \ourname achieves a \fscore of 0.97. This shows the advantage of the triplet-loss function of \ourname, as here the triplet loss provides the training process of the Siamese network with better choices for negative samples, leading to a more powerful embedding extraction.

\subsubsection{Individual modality performance}
\label{sec:modalityperformance}
\ourname performs an implicit fusion of the individual sensors' values, i.e., the raw data of multiple sensors (modalities) is consolidated by the CNN before the model extracts any information. This stands in contrast to classical fusion where the data of each sensor is processed separately (i.e., by separate NNs) and the results of the complete processing pipeline are combined at the end. In \sect\ref{sec:eval-scorefusion} we compare the implicit fusion of \ourname against different standard fusion techniques, e.g., summing up or using a Multi-Layer-Perceptron (MLP).
In order to assess the performance of each modality individually, in the following first we train \ourname using the data of every sensor. Therefore, we obtain three models trained on three modalities (i.e., accelerometer, gyroscope, and magnetometer). For this assessment, the Siamese Network is trained with 100 epochs and a batch size of 1024 while the corresponding decision network is trained with 10 epochs and a batch size of 64.

\begin{figure}
	\centering
	\includegraphics[width=0.48\textwidth]{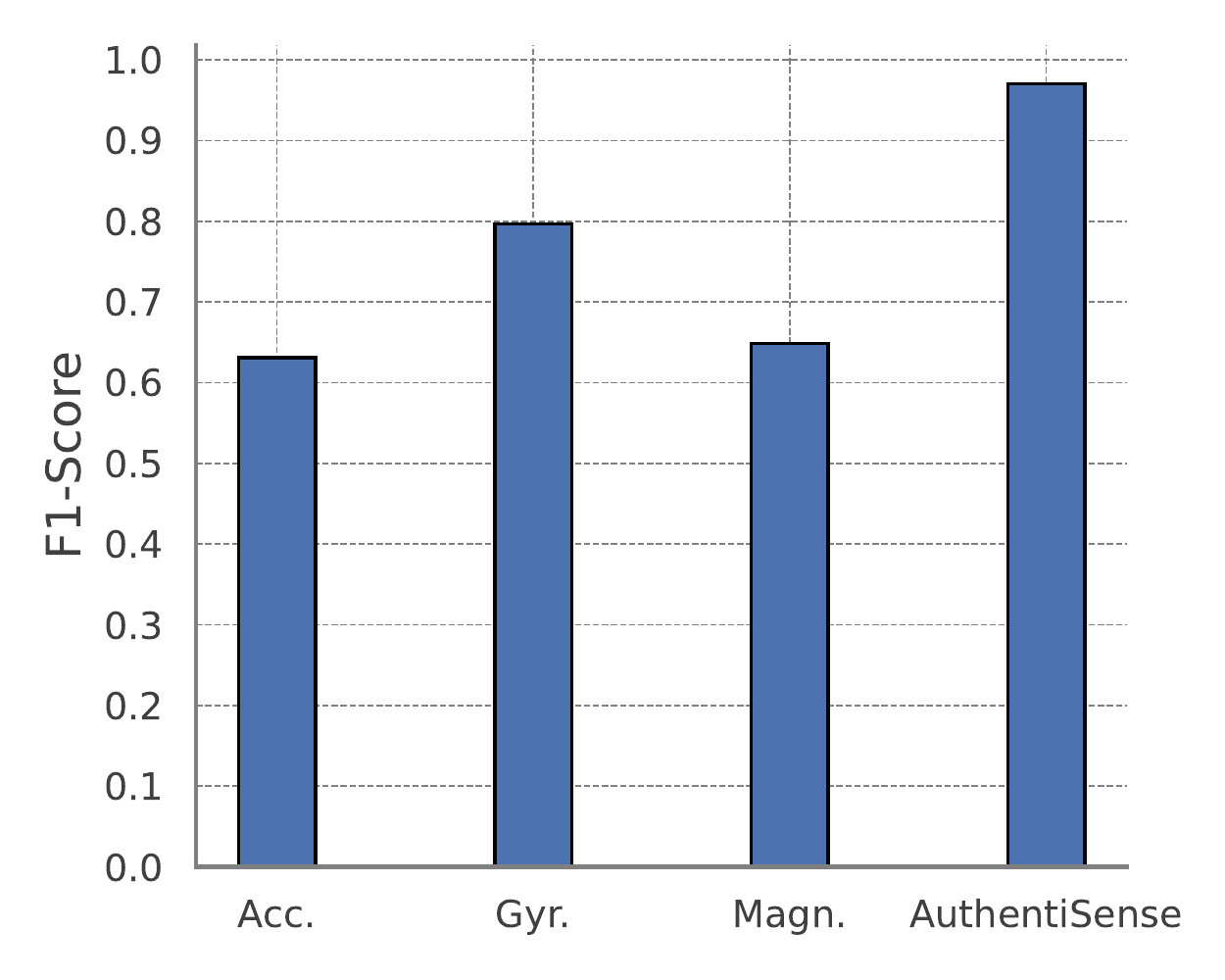}
	\vspace{-2.5em}
	\caption{The \fscores of single-modality-trained models and of \ourname.}
	\label{fig:modality-performance}
\end{figure}

The results are shown in Fig.~\ref{fig:modality-performance} and Tab.~\ref{tab:testdataseteers} where they are also compared with the performance of \ourname.
Both accelerometer- and magnetometer-trained models, although each performing good on the train samples, suffer from a sharp performance decline when evaluated on the test samples. In both cases, the precision remains mostly unchanged above 0.90, while the recall drops by twenty to thirty percentage points (from 0.89 to 0.49 in the case of the magnetometer).

The gyroscope-trained model, on the other hand, offers the best performance (\fscore: 0.80) with relatively stable precision and recall values (0.76 and 0.83 for test samples). 
Nevertheless, even the gyroscope-trained model is not comparable with the performance of \ourname (\fscore: 0.97), as shown in Fig.~\ref{fig:modality-performance}.

\begin{table}[tb]
\centering

        \centering
        \begin{tabular}{ll|r|r|r|r|c|}
            \cline{3-7}
            &    & \multicolumn{5}{c|}{Authentication window length (Sec.)}                                                                    \\ \cline{3-7} 
                                                          &    & \multicolumn{1}{c|}{1} & \multicolumn{1}{c|}{3} & \multicolumn{1}{c|}{5} & \multicolumn{1}{c|}{10} & \multicolumn{1}{c|}{15} \\ \hline
            \multicolumn{1}{|l|}{\multirow{5}{*}{\STAB{\rotatebox[origin=c]{90}{\nComparison}}}} & 1   & \multicolumn{1}{l|}{0.92} & \multicolumn{1}{l|}{0.91} & \multicolumn{1}{l|}{0.90} &\multicolumn{1}{l|}{0.91} & 0.89\\ \cline{2-7}
\multicolumn{1}{|l|}{} & 2 & \multicolumn{1}{l|}{0.92} & \multicolumn{1}{l|}{0.91} & \multicolumn{1}{l|}{0.90} &\multicolumn{1}{l|}{0.91} & 0.88\\ \cline{2-7}
\multicolumn{1}{|l|}{} & 3 & \multicolumn{1}{l|}{0.92} & \multicolumn{1}{l|}{0.91} & \multicolumn{1}{l|}{0.89} &\multicolumn{1}{l|}{0.91} & 0.86\\ \cline{2-7}
\multicolumn{1}{|l|}{} & 4 & \multicolumn{1}{l|}{0.92} & \multicolumn{1}{l|}{0.90} & \multicolumn{1}{l|}{0.92} &\multicolumn{1}{l|}{0.90} & 0.86\\ \cline{2-7}
\multicolumn{1}{|l|}{} & 5 & \multicolumn{1}{l|}{0.93} & \multicolumn{1}{l|}{0.90} & \multicolumn{1}{l|}{0.91} &\multicolumn{1}{l|}{0.90} & 0.86\\ \hline
        \end{tabular}

    \caption{\fscore for pair training with the contrastive loss.}
    \label{tab:user-based-auth}
\end{table}

It must be noted that an optimal \fscore does not imply an optimal EER (where FAR=FRR), since the threshold for the highest score is sometimes located at a position, where FAR$\neq$FRR. This is also the case in this experiment, as found in Tab.~\ref{tab:testdataseteers}. Here, it is the accelerometer-trained model that achieves the best results in regard to EER. Still, the achieved EER of 0.1706 is more than four times worse than the corresponding EER of \ourname. Overall, none of the single-modality-trained models is able to compete with \ourname, affirming the benefits of an architecture that feeds the consolidated data of all motion sensors into a single neural network. 

\begin{table}[h]
    \centering
    \begin{tabular}{r|cccc|}
    \cline{2-5}
               &            \multicolumn{3}{c|}{Modality} & \multicolumn{1}{c|}{\multirow{2}*{\ourname}} \\ \cline{1-4}
    \multicolumn{1}{|r|}{Samples} & \multicolumn{1}{c|}{Acc.} & \multicolumn{1}{c|}{Gyr.} & \multicolumn{1}{c|}{Mag.} & \multicolumn{1}{c|}{}\\ \hline

    \multicolumn{1}{|r|}{Test} & \multicolumn{1}{c|}{0.1706} & \multicolumn{1}{c|}{0.2308} & \multicolumn{1}{c|}{0.1902} & \multicolumn{1}{c|}{0.0371}\\ \hline
    \end{tabular}
    \caption{EERs of the test set of each single-modality trained model and of \ourname.}
    \label{tab:testdataseteers}
\end{table}

\subsubsection{Fusion of modalities}
\label{sec:eval-scorefusion}
Single-modality-trained models do not perform well on their own, as found in Section~\ref{sec:modalityperformance}. With the fusion of their matching scores, an additional performance improvement can be attempted. In the following, we evaluate the effectiveness of such score level fusions. The performance of fusion is compared both with the best performing single-modality-trained models (described in \sect~\ref{sec:modalityperformance}) and with the best performing configuration of \ourname.\par

We evaluate multiple fusion methods. For each method,
different matching score normalization techniques are applied. The individual methods are described in App.~\ref{sec:appendix:scorefusion} in detail.
For training the fusion and normalization methods, we use the scores that were predicted for the 50,000 training samples in \sect~\ref{sec:modalityperformance}.
For evaluation, the trained fusion methods are applied to the matching scores of the corresponding 10,000 test samples. The results of all utilized fusion techniques in regard to their best performing normalization method are described in Tab.~\ref{tab:fusionresults}.

Measured by EER, the best results are obtained with EER-weighted sum fusion with min-max normalization (EER: 0.1419).
Compared with the best single-modality-trained model (EER: 0.1706; cf. Tab.~\ref{tab:testdataseteers}), this experiment shows that score level fusion can offer a performance boost over single-modality based systems in regard to the EER. Such boost, however, is not achieved for the \fscore. The best results, measured by \fscore, are obtained with Z-Score normalized MLP Fusion. This technique's \fscore surpasses the scores of the accelerometer and magnetometer trained models. However, it is slightly below the score of the gyroscope trained model (cf. Fig.~\ref{fig:modality-performance}). In general, all fusion methods reveal high precision and low recall values, as very similarly experienced with the accelerometer and magnetometer trained single-modality models in \sect\ref{sec:modalityperformance}. Overall, the improvement of score level fusion over single-modality-trained models is only marginal.

The performance of \ourname (\fscore: 0.97; EER: 0.0371) is not reached by any of the single-modality-trained models (best \fscore: 0.80; best EER: 0.1706). With only marginal improvement of score level fusion over the latter (best \fscore: 0.76; best EER: 0.1419), score level fusion is equally incapable of achieving significantly better results. It must therefore be acknowledged that, for the use cases discussed in this work, our proposed sensor level fusion approach of \ourname constitutes a substantial advantage over score level fusion.

\begin{table}
\centering

\begin{tabular}{l|ccc|}

\hline
\multicolumn{1}{|l|}{\textbf{Fusion}} & \multicolumn{1}{c|}{\textbf{Normalization}} & \multicolumn{1}{c|}{\textbf{\fscore}} & \multicolumn{1}{c|}{\textbf{EER}} \\ \hline
\multicolumn{1}{|l|}{Linear Reg.} & \multicolumn{1}{c|}{Not normalized} & \multicolumn{1}{c|}{0.69} & \multicolumn{1}{c|}{0.1454}  \\ \hline
\multicolumn{1}{|l|}{Logistic Reg.} & \multicolumn{1}{c|}{Not normalized} & \multicolumn{1}{c|}{0.69} & \multicolumn{1}{c|}{0.1453}  \\ \hline
\multicolumn{1}{|l|}{MLP} & \multicolumn{1}{c|}{Z-Score} & \multicolumn{1}{c|}{0.76} & \multicolumn{1}{c|}{0.1451}  \\ \hline
\multicolumn{1}{|l|}{Sum} & \multicolumn{1}{c|}{Min-Max} & \multicolumn{1}{c|}{0.75} & \multicolumn{1}{c|}{0.1532}  \\ \hline
\multicolumn{1}{|l|}{EER-W. Sum} & \multicolumn{1}{c|}{Min-Max} & \multicolumn{1}{c|}{0.67} & \multicolumn{1}{c|}{0.1419} \\ \hline \hline
\multicolumn{2}{|l|}{\ourname} & 0.97 & \multicolumn{1}{|c|}{0.0371}\\\hline

\end{tabular}

\caption{Performance of all utilized fusion methods. For each fusion method, only the best performing normalization method is listed.}
\label{tab:fusionresults}
\end{table}

\section{Related Work}
\label{sec:related}

In the past, different approaches for authenticating the users of mobile devices have been developed. These approaches use either statistical features extracted on touch gestures~\cite{lu2015safeguard,xu2014towards,frank2012touchalytics,zhao2014mobile,zhao2013continuous,islam2021scalable,chen2020listen}, values that are captured by the sensors for identifying the user based on its motions~\cite{li2020scanet,hu2018cnnauth,buriro2017behavioral,buriro2017please,amini2018deepauth,zhu2013sensec,lee2017secure,haring2018pick,benegui2020convolutional,ehatisham2018continuous,lee2017implicit,wang2021framework,CNN-based}, or correlate the touch gestures and motion sensors~\cite{deb2019actions, bo2013silentsense,buriro2016hold,buriro2015touchstroke}.
However, these existing approaches are 1) restricted to identifying a user among a set of known users, thus can only detect intruders that were part of the training data~\cite{ehatisham2018continuous,frank2012touchalytics,lu2015safeguard,wang2021framework,xu2014towards}, 2) require training the model when a new user joins the system limiting their scalability~\cite{benegui2020convolutional,Centeno2018,zhu2013sensec}, or 3) work only in certain situations and not in a continuous way, e.g., when picking up the phone~\cite{buriro2019answerauth,buriro2016hold,buriro2015touchstroke}. Compared to the existing works, by combining different data sources (accelerometer, gyroscope, magnetometer) with a few-shot learning approach, \ourname can authenticate the user reliably, while requiring only a minimum of user interaction. In the following, we discuss the existing approaches in more detail and compare them with \ourname.

\subsection{Touch and Typing Behavior}
Lu \etal~\cite{lu2015safeguard}, Xu \etal~\cite{xu2014towards}, and Frank \etal~\cite{frank2012touchalytics} collect touch-event features such as the position (X and Y coordinates), pressure and gesture velocity for training a Support-Vector-Machine (SVM).  Lu \etal~\cite{lu2015safeguard} and Xu \etal~\cite{xu2014towards} use training data from the benign user and another user that they define as the attacker for training the SVM to distinguish them. 
The SVM Frank \etal is trained for recognizing the user out of a set of known users~\cite{frank2012touchalytics}. However, these approaches are not practical, since they cannot detect unknown attackers but only distinguish between known persons. Therefore, they cannot reliably prevent unknown users from accessing the system as the respective scheme will recognize unknown intruders as one of the users from training, whose behavior is most similar to the intruders' behavior. \ourname addresses this issue by using a few-shot learning approach that detects arbitrary intruders even if no samples of the benign user were part of the training data.

Zhao \etal convert touch features (position, pressure etc.) into an image, called Graphic Touch Gesture Feature (GTGF)~\cite{zhao2013continuous}. An extension uses a statistical feature model on the features to make the system more robust against variations in the benign behavior~\cite{zhao2014mobile}. However, these approaches are limited since they require the user to perform specific gestures on the phone for authentication, e.g., swiping from the top-right corner to the bottom-left corner. In contrast, \ourname continuously analyses the sensor values for identifying the user without requesting any specific gesture patterns.

Chen \etal combine features that are extracted from the touch gestures with acoustical sensor data~\cite{chen2020listen}. However, their similarity technique requires a comprehensive set of comparison samples, while for \ourname less than 5 shots are sufficient.  Similarly, also the approach of Karanikiotis \etal requires a high amount of training data for each user to extract standard features from the touch gestures, like the swiping duration~\cite{karanikiotis2020continuous} and apply a SVM. 

The solution by Islam \etal requires the users to solve a challenge, e.g., draw a circle~\cite{islam2021scalable}. However, with this explicit authentication action, there is no advantage compared to standard explicit authentication approaches, e.g., scanning the fingerprint. If this is performed continuously, such a scheme is likely to disturb the user, causing the user to turn-off the authentication scheme. In comparison, \ourname can run unobtrusively in the background and becomes only visible, when the user is not recognized and the device is looked.

Further approaches analyze the mobile keystroke dynamics~\cite{inguanez2016securing,gurary2016implicit,buschek2018researchime,acien2019keystroke,keystroke-2017,knn-one-model}. 
However, keystroke dynamics are not well-suitable for continuous authentication on mobile platforms, as mobile apps do not frequently involve keyboard-based user interactions.
In contrast, \ourname utilizes only motion patterns (sensor values) and can  frequently and reliably authenticate users.

\subsection{Motion-based Approaches}

In order to obtain a more augmented set of machine learning features, the CNN-based approach SCANet~\cite{li2020scanet} and its predecessor CNNAuth~\cite{hu2018cnnauth} perform a transformation of the temporal sensor data into the frequency domain and utilize a CNN that is trained to recognize the legitimate user. 
However, since both approaches train CNNs for each user individually, they require a large amount of training data 
during the system enrollment. In comparison, the few-shot approach of \ourname allows an effective authentication of the user and requires only very few user samples, e.g., just a single sample of the current user for comparing the input without the need for subsequent training which would require hundreds of samples.

Buriro \etal proposed an approach that uses the sensor data of the mobile phone for authenticating the user after unlocking the phone, assuming that the movements in this scenario are always similar for the same user~\cite{buriro2017please}. In comparison to \ourname that performs continuous authentication while the phone is being used, the approach of Buriro \etal only authenticates the user once, i.e., right after the phone is unlocked, s.t. their approach is orthogonal to \ourname.

ActiveAuth uses a Gaussian Data Description verifier for identifying the user~\cite{buriro2017behavioral}. Analogously to the previous approach, also ActiveAuth does not perform continuous authentication but authenticates the user only in certain situations, e.g., when uninstalling an application.

DeepAuth applies a user-specific RNN on data from the accelerometer and gyroscope~\cite{amini2018deepauth}. By using Long-Short-Term-Memories (LSTMs), the system can effectively process time-series data~\cite{amini2018deepauth}. However, for training the user-dependent RNN, DeepAuth first needs to collect a sufficient amount of data, while \ourname just requires few comparison samples, making \ourname more practical.

Zhu \etal use an n-gram Markov model based on data from accelerometer, gyroscope and magnetometer to authenticate the user~\cite{zhu2013sensec}. In contrast to the user-agnostic approach of \ourname, their model is user-specific and requires training efforts for each new user. 

Lee \etal proposed an approach that identifies users based on the motion that occurs while their phone is being picked up from a surface~\cite{lee2017secure}. 
A similar system is also introduced by Haring \etal~\cite{haring2018pick}. Compared to \ourname, their approach is is only effective in a single use-case, i.e., when a user picks up the phone. Therefore, it cannot provide continuous nor passive authentication while \ourname provides both authentication types.

Discrete motion is also generated when a user taps the phone's touchscreen. The authors of~\cite{benegui2020convolutional} capture the motion sensor data of such events and feed a CNN model with it. An SVM is trained on the generated CNN features as binary classifier using data from the legitimate user and a non-legitimate user~\cite{benegui2020convolutional}. However, because the SVM is trained only to distinguish the legitimate user and the data from other people that was used during training, it cannot detect an intruder whose behavior was not covered by the training data and is, e.g., more similar to the benign user than to the non-legitimate user that was used during training.

The authors of~\cite{lee2017implicit} propose an authentication system that enriches the motion data of phones with additional motion information obtained from user-owned wearable devices such as Smartwatches.
While this system does not only require the presence of a wearable device (the performance is considerably degraded without such auxiliary information), it also requires an available cloud to which the user's motion data must be sent for training. In contrast, \ourname requires neither subsequent training nor additional devices or cloud services.

In order to authenticate users continuously over a longer period, Ehatisham-ul-Haq \etal utilize a phone's motion sensors to detect different types of human activity and identify users based on their everyday activity patterns using an SVM~\cite{ehatisham2018continuous}. Wang \etal use an accelerometer to identify the user out of a set of known users, also using a Siamese network \cite{wang2021framework}. However, their approach is not suitable for authentication as it cannot distinguish between the benign user and unknown users which were not present in the training data. Centeno \etal train a One-Class SVM to identify the benign user after extracting features from sensor data using a Siamese network~\cite{CNN-based}. However, their approach does not scale, because for training the One-class SVM, a high number of training samples for each user is needed, while \ourname requires only few samples for comparison.

\subsection{Touch-Motion Behavior}
Deb \etal \cite{deb2019actions} combine raw horizontal and vertical scrolling logs for touch behaviour, and Fourier-transformed  accelerometer, gyroscope or magnetometer data. For each modality, they train a separate LSTM network using contrastive loss, and modalities are combined with score level fusion. However, they only train a distance threshold, which, given the high variance of behavior data, does not solve the scalability problem.

Buriro \etal proposed a sensor-augmented authentication model during for PIN entry under different user positions, using summary statistic representations for sensor data and inter-key latency for fixed-length PIN entries, fed into binary classifiers~\cite{buriro2015touchstroke}. However, since they analyze the user-behavior during the PIN insertion their approach focuses on improving the security of classical authentication methods. In comparison, \ourname performs a continuous authentication allowing to detect intruders also after the regular authentication. Therefore, the approach of Buriro \etal is orthogonal to \ourname.

In a follow up work, Buriro \etal \cite{buriro2016hold} used a one-class Multilayer Perceptron on a behavioural dataset of 30 users signing on their touch screens along with sensor data. For each user, a Multilayer Perceptron was trained with owner data only, without any impostor samples. However, analogously to the previous work, also this approach cannot be applied continuously. AnswerAuth analyses the phones' sensor data while the user is performing certain actions, e.g., like sliding or lifting the phone. AnswerAuth applies a Random-forest classifier to identify the current user out of a set of known users~\cite{buriro2019answerauth}. However, since AnswerAuth can only identify a person from a set of known users, for which training samples were recorded, it cannot identify intruders.

Multiple approaches train models for each user individually for authenticating them based on touch-gestures and sensor data. Incel~\etal use sensor and touch data to authenticate users in a banking app. They trained binary SVMs with RBF kernels to authenticate users~\cite{DAKOTA}. Humayoun \etal train a DNN to combine features that are extracted from touch-gestures with sensor data~\cite{humayoun2022touch}. Abuhamad \etal~\cite{autosen} utilize LSTM Networks to authenticate users with short authentication windows at a high frequency. Acien \etal use touch, accelerometer, gyroscope, keystroke, WiFi, GPS, and App Usage data to authenticate user. They profiled users by training separate RBF-SVM classifiers for each user and each modality, ending up with seven models for every single user, combined with score-level fusion~\cite{multilock}.
However, since these approaches train separate models for each user, they need to collect a high number of training samples during the enrollment phase. Since normal users are unlikely to perform certain gestures for many times in during the setup, these approaches are not scalable.

To summarize, \ourname utilizes only standard built-in sensors (i.e., accelerometer, gyroscope, magnetometer) for authentication purpose. These sensors are typically available on mobile devices, therefore, \ourname can work in most settings. Furthermore, \ourname is trained in a user-agnostic fashion and even works for users it is not trained for. It is scalable and allows users to be on-boarded with only a small number of enrollment samples without any further training. The use of a feature extractor network also makes it possible to enroll users by only storing behavioral feature vectors rather than raw behavior data, which makes it less privacy invasive and reduces the total authentication overhead.

\section{Discussion}
\label{sec:discussion}
Behavior biometrics systems are becoming more widespread and effective as technology advances, they are not a bullet-proof solution for authentication or identification. In the following, we outline some of the limitations of behavioral biometrics authentication systems including \ourname.

Behavioral biometrics authentication systems may suffer from failure to enrol. This happens when a reference sample for biometrics cannot be successfully created at the time of enrolment due to a number of factors, such as low-quality sensors, poor environmental conditions, physical or medical conditions of the individuals~\cite{ovic}. Ensuring effective enrolment is crucial to the successful operation of a biometrics authentication system.

The use of behavioral biometrics, as with other security measures, has vulnerabilities and can be compromised. The sensor data can be spoofed or retrieved through side-channel attacks. However, behavioral biometrics spoofing can be mitigated by hardware security extension technologies or through sensor data obfuscation.

Another limitation of behavioral biometrics systems is that unlike traditional authentication methods (i.e., passwords), behavioral biometric characteristics cannot be reissued or cancelled. In case of being compromised, if not impossible it can be extremely difficult to change the characteristics. This makes it problematic when using those behavioral biometric characteristic for future authentication.

The matching of an individual with a reference information stored in the system is a probabilistic calculation. There are errors (i.e., false acceptance and rejection rates) that may be influenced by a range of factors. The user interaction with a sensor may differ between the enrollment and recognition phases or, in rare cases, individuals may share similar behavioral biometric characteristics. In addition, factors such as aging, or medical conditions can also affect individuals' behavioral biometric characteristic between the enrollment and recognition stages.

Behavioral biometrics like many other technologies can pose challenges to privacy.  Since in behavioral biometrics authentication systems the information collection is covert or passive, individuals may be unable to provide consent or control over what information is collected or how it is used.  Another privacy risk is depending on the characteristics, some behavioral biometrics (i.e., motion patterns) could potentially reveal secondary information (i.e., health-related issues) about an individual who may not want to provide that information.

Furthermore, during training of AuthentiSense, our loss function does not take the context into account, and triplets are formed using only ``subject'' information, where it is possible to encounter a positive sample from a very different context than the anchor. For instance, in the anchor, the user could be on one device and scrolling, and in the positive sample, the user could be typing on a different device. In this case, it would be difficult for the network to learn a function that can recognize the similarities between the two samples.

In future work, a triplet mining procedure can be implemented to consider more than label information when selecting positive and negative samples, such as posture, device model, or any other relevant \emph{contextual} information which could affect the behavior samples collected. We hypothesize that a feature extractor model may create more compact clusters for different devices
or postures, if positive samples are only drawn for the same context, rather than one big loose cluster for all of a given user's behavior samples.

\section{Conclusion}
\label{sec:conclusion}

We propose \ourname, a user-agnostic behavioral biometrics authentication system that continuously uses motion patterns while interacting with mobile apps and regularly validates the authenticity of a user after the user has logged in. The passive nature of \ourname makes it non-intrusive to the users' experience and takes the burden off the users, offering a frictionless and quick authentication method. We utilize a few-shot learning-based model called Siamese network and train an efficient model that is not user-specific. Our proposed approach is highly scalable and fast. It does not require hand-crafted features for model training and does not need to be re-trained when users are dynamically changing (i.e., joining or leaving the system). When evaluated in a Few-shot fashion, our system needs only a few behavior samples per user. We conduct an extensive and systematic measurement study and analyze the impact of different parameters such as choice of loss functions, size of the authentication window time, and n-shots. Our evaluation results demonstrate that \ourname can achieve an accuracy of 97\% in terms of F1-score for 3-shot verification and can accurately authenticate users already after 1 second of user interaction.

\section*{Acknowledgment}
We would like to thank Intel Private AI center and BMBF for their support of this research.
{\footnotesize \bibliographystyle{acm}
\bibliography{ref}}
\appendix
\section{Appendix}
\label{sec:appendix}

\subsection{Contrastive loss}
\label{sec:appendixA}
In contrastive loss, two data samples are fed into the Siamese networks one after the other to get embedding vectors. Then in the latent embedding space, the distance $D$ between the two embedding vectors are computed. Finally, the calculated distance $D$ is substituted into the loss function (Eq.~\ref{eq:cont}) and the Siamese network is trained via backpropagation for better latent vector embedding. The loss function is defined as below:

\begin{equation} \label{eq:cont}
   \mathcal{L}(Y,D) = (Y)\cdot(D^{2}) + (1-Y) \cdot{max(margin-D,0)}^{2}
\end{equation}

Where the $Y$ value is ground truth label. $Y=1$ if the input pairs are of the same user, and $Y=0$ otherwise. The max function takes the largest value of 0 and the margin (a pre-defined hyper-parameter) minus the distance.

\subsection{Neural Network}
\label{app:nn}

Neural Networks use a so-called training dataset to learn a specific task. They take samples from a domain $\mathcal{D}$, e.g., images or sensor values, as input and predict an output vector $l$ indicating e.g., the recognized category, from a target set $\mathcal{L}$. The parameters of a NN, therefore, realize a function $f:\mathcal{D}\mapsto\mathcal{L}$. It should be noted that $l$ can also be a binary label, then $\mathcal{L}=\{0,1\}$. 

NNs consist of individual neurons, which are grouped into layers. The input of the NN is given to first layer, that determines for each neuron an activation status. The activation status of each neuron of the current layer is then given to the neurons of the next layer. The activation $n_{l,i}$ of a (linear) neuron $i$ in layer $l$, which uses the activations of $K$ neurons from the previous layer as input is then given by:
\begin{equation}
    n_{l,i} = g(b_{l,i} + \sum\limits_{k=0}^{K} w_{l,i,k} \cdot n_{l-1,k})
\end{equation}
where $b_{l,i}$ and $w_{l,i,k}$ are parameters of the NN that are optimized during the training phase for the respective task. $g$ is the so-called activation function, a non-linear function as $\text{relu}(x) = \max(0, x)$ that is used to determine activation of the current neuron and allows the NN to recognize also non-linear patterns~\cite{goodfellow2016deep}.

\subsection{1-dimensional Convolutional Neural Network.} 
\label{app:cnn1d}
In Convolutional Neural Networks (CNNs) 
each neuron only uses a small window of neighboring inputs to determine its activation status, e.g., neighboring time steps for sequential data~\cite{goodfellow2016deep}. CNNs are taking their name from the mathematical tool called convolution,

which is a linear operator used for feature extraction where a small 
filter or a kernel, is slided across an input sequence. Iterating on the input piece by piece allows the model to extract small features producing in output a feature map, that depends on both the input values and the kernel weights. Different kernels can therefore be thought of as different feature extractors.

Although CNNs are most popularly used for two-dimensional inputs like images, they can be generalized to other dimensions, and one-dimensional CNNs (or CNN1D) are more suitable for certain applications dealing with sequential data like one-dimensional sensor readings~\cite{KIRANYAZ2021107398}.

Furthermore, another advantage of CNN1D is that they have less trainable parameters (i.e., weights), therefore, they are less complex and faster to train compared to 2D-CNNs allowing smaller sets of training data, making them more suitable for real-time and low-cost applications \cite{KIRANYAZ2021107398}.

\subsection{Score Fusion}
\label{sec:appendix:scorefusion}

In the field of behavioral biometrics, information from different sensors is not always processed by a single classifier (sensor level fusion). Instead, multiple classifiers (matchers) are typically used that each produce a matching score. To form a final classification, a vector of matching scores $s=[s_1,\dots,s_R]$, obtained from $R$ matchers, is fused using special techniques (e.g., score level fusion) in order to classify samples as either \textit{genuine} or \textit{impostor}~\cite{ross2006handbook}. Having a broader collection of information available, score fusion models can potentially achieve better performance than each of the individual matchers does on their own.
In the following, we describe a selection of fusion and normalization techniques.

\subsubsection{Normalization Techniques}
simple rule-based fusion techniques require matching scores to lie in a common range. This is achieved through normalization methods. Each method has different impact on the fusion technique's performance.

\noindent \textit{The Z-Score Normalization:} is a method that normalizes the $j^{th}$ matcher by utilizing arithmetic mean ($\mu_j$) and standard deviation ($\sigma_j$) of the available match scores. The normalized value of the $i^{th}$ match score $s_j^i$, as specified in~\cite{ross2006handbook}, is

\begin{equation}
ns_j^i=\dfrac{s_j^i-\mu_j}{\sigma_j}.
\end{equation}

where feature distributions are centered around zero with unit standard deviation.

\noindent \textit{Min-Max Normalization:} Min-Max normalization maps values between 0 and 1. Just as z-score normalization, this method is not robust, because outliers can be mapped to the outside of the lower and upper boundaries. Let $s_j^i$, as defined in~\cite{ross2006handbook}, be the $j^{th}$ matcher's $i^{th}$ matching score of the $N$ scores that are used for tuning the normalizer. Then the normalized value of the $i^{th}$ match score $s_j^i$ is

\begin{equation}
ns_j^i=\dfrac{s_j^i-\min^N_{k=1} s_j^k}{\max^N_{k=1} s_j^k - \min^N_{k=1} s_j^k}.
\end{equation}

where 0 corresponds to the minimum observation and 1 corresponds to the maximum

\subsubsection{Fusion Techniques} Fusion techniques include (i) weighted sum fusion, (ii) EER-weighted sum fusion, (iii) linear and logistic regression fusion as well as (iv) multi-layer perceptron (MLP) methods. 

For each technique, matching scores are optionally normalized.

\noindent \textit{Weighted Sum Fusion:} with this fusion technique, the weighted matching scores of all matchers are added up. For a binary classification problem, it is sufficient to focus on just one class (\textit{genuine}) and determine, if the sum passes a certain threshold.

The fused score of the normalized $i^{th}$ match score is defined as
\begin{equation}
    f^i=\sum^R_{j=1}w_jns^i_j
    \label{eq:wsumfusion}
\end{equation} where $w_j$ is the $j^{th}$ matcher's weight~\cite{alsaade2008score}. For pure sum fusion, the weight of all matchers is set to $1$.\par

It is necessary to find an optimal threshold $t$ in order to perform final classification on the produced fusion score $f^i$ (\textit{genuine}, if $f^i\geq t$; \textit{impostor} else). This optimization is performed during fusion training and utilizes a sigmoid activated linear regression model that takes single fusion scores as input. Such model is described below.

\noindent \textit{EER-Weighted Sum Fusion:} it sets the weights of accurate matchers higher than those of less accurate ones. The weight $w_j$ of the $j^{th}$ matcher is defined as
\begin{equation}
    w_j = \frac{1}{EER_j\left(\sum_{i=k}^R\frac{1}{EER_k}\right)}
\end{equation} where $EER_j$ represents the $j^{th}$ matcher's equal error rate~\cite{alsaade2008score}. All weights are then applied on the weighted sum fusion defined in Equation~\ref{eq:wsumfusion}.

\noindent \textit{Linear and Logistic Regression Fusion:} both regression fusion techniques are a form of classifier-based score fusion where a boundary between the classes \textit{genuine} and \textit{impostor} is learned~\cite{ross2006handbook}. For linear regression, the following function is approximated
\begin{equation}
    f^i = \beta + \sum^R_{j=1}w_j\cdot ns^i_j,
\end{equation}
while for logistic regression,
\begin{equation}
    f^i = \frac{e^{\beta + \sum^R_{j=1}w_j\cdot ns^i_j}}{1+e^{\beta + \sum^R_{j=1}w_j\cdot ns^i_j}}
\end{equation} is approximated. The produced value $f^i$ is then passed to a sigmoid function for classification. For optimization, the machine learning model adjusts both the bias $\beta$ and each $j^{th}$ matcher's weight $w_j$.

\noindent \textit{Multi-Layer Perceptron (MLP) Fusion:} MLPs are fully connected neural networks with one or more hidden layers. As defined in~\cite{alsaade2008score}, MLP fusion with one hidden layer applies the function
\begin{equation}
    f^i = s_o\left( \sum^{M_H}_{j=0} w_{ji} s_h\left( \sum^{M_I}_{k=0} w_{ki} ns^i \right) \right)
\end{equation} on the $i^{th}$ normalized score $ns^i$ to produce the $i^{th}$ fused score $f^i$. The sigmoid activation functions of the output ($s_o$) and the hidden layer ($s_h$) are provided with the sum of $M_H$ weighted ($w_{ji}$) hidden and $M_I$ weighted ($w_{ki}$) input nodes respectively, whose weights are optimized during the training process~\cite{alsaade2008score}.




%

\end{document}